\documentclass[twocolumn]{aastex62}
\usepackage{xspace}
\usepackage{multirow}

\newcommand{\teq}{\ensuremath{T_\mathrm{eq}}\xspace}
\newcommand{\tint}{\ensuremath{T_\mathrm{int}}\xspace}

\newcommand{\kzz}{\ensuremath{K_{zz}}}
\newcommand{\soo}{SO$_2$\xspace}
\newcommand{\coo}{CO$_2$\xspace}
\newcommand{\water}{H$_2$O\xspace}
\newcommand{\hhs}{H$_2$S\xspace}
\newcommand{\methane}{CH$_4$\xspace}

\newcommand{\referee}{} 

\newenvironment{my_itemize}{
\begin{itemize}
  \setlength{\itemsep}{1pt}
  \setlength{\parskip}{0pt}
  \setlength{\parsep}{0pt}}{\end{itemize}
}

\shorttitle{The \soo\ Shoreline}
\shortauthors{Crossfield, et al.}

\begin{document}

\title{Mapping the \soo shoreline in gas giant exoplanets}

\correspondingauthor{Ian J.\ M.\ Crossfield}
\email{ianc@ku.edu}

\author{Ian J.\ M.\ Crossfield}
\affiliation{Department of Physics and Astronomy, University of Kansas, Lawrence, KS, USA}
\affiliation{Max-Planck-Institut f\"ur Astronomie, K\"onigstuhl 17, D-69117 Heidelberg, Germany}

\author[0000-0003-0973-8426]{Eva-Maria Ahrer}
\affiliation{Max-Planck-Institut f\"ur Astronomie, K\"onigstuhl 17, D-69117 Heidelberg, Germany}

\author[0000-0002-2072-6541]{Jonathan Brande}
\affiliation{Department of Physics and Astronomy, University of Kansas, Lawrence, KS, USA}

\author[0000-0003-0514-1147]{Laura Kreidberg}
\affiliation{Max-Planck-Institut f\"ur Astronomie, K\"onigstuhl 17, D-69117 Heidelberg, Germany}

\author[0000-0003-3667-8633]{Joshua Lothringer}
\affiliation{Space Telescope Science Institute, Baltimore, MD 21218, USA}

\author[0000-0002-2875-917X]{Caroline Piaulet-Ghorayeb}
\altaffiliation{E. Margaret Burbridge Postdoctoral Fellow}
\affiliation{Department of Astronomy \& Astrophysics, University of Chicago, 5640 South Ellis Avenue, Chicago, IL 60637, USA}

\author[0009-0009-8749-9513]{Jesse Polman} %
\affiliation{Division of Space Research and Planetary Sciences, Physics Institute, University of Bern, Gesellschaftsstrasse 6, 3012 Bern, Switzerland}

\author[0000-0003-0156-4564]{Luis Welbanks}
\affiliation{School of Earth and Space Exploration, Arizona
State University, Tempe, AZ, USA}

\author[0000-0002-4207-6615]{James Kirk}
\affiliation{Department of Physics, Imperial College London, Prince Consort Road, London SW7 2AZ, UK}

\author[0000-0002-4250-0957]{Diana Powell}
\affiliation{Department of Astronomy \& Astrophysics, University of Chicago, Chicago, IL 60637, USA}

\author[0000-0001-8427-9173]{Niloofar Khorshid}
\affiliation{Department of Space, Earth and Environment, Chalmers University of Technology, Gothenburg, 412 96, Sweden}

\begin{abstract}
JWST has revealed sulfur chemistry in giant exoplanet atmospheres, where molecules such as SO2 trace photochemistry, metallicity, and formation and migration. To ascertain the conditions that determine whether (or how much) \soo, \hhs, and other sulfur-bearing species are present in exoplanet atmospheres, we present a grid of planetary atmospheres covering metallicities from 0.3--1000$\times$ Solar and temperatures from 250--2050 K. These models map out the `\soo\ shoreline,' the region of metallicity and irradiation for which \soo\ may be sufficiently abundant to be detectable. \soo\ is a sensitive indicator of metallicity; expected \soo\ abundances also depend strongly on overall temperature and C/O ratio; the \soo\ abundance depends surprisingly weakly on XUV irradiation, also weakly on \kzz\ (for $T_\mathrm{eq} \gtrsim 600$\,K), and is essentially independent of internal temperature. Despite its detection in a growing number of giant planets, \soo\ is never the dominant sulfur-bearing molecule: depending on temperature and metallicity, H$_2$S, S$_2$, NS, SO, SH, and even S$_8$ or atomic S are frequently as common (or more so) as \soo. Nonetheless \soo\ remains the most easily detectable sulfur-bearing species, followed by \hhs, though perhaps SO and SH could be detectable in some gas giants. Aside from a pressing need for additional observational constraints on sulfur, we also identify the need for future work to account for the effects of clouds and hazes, fully self-consistent atmospheric models, 2D and 3D models, a wider range of planetary masses and radii, and studies to measure and refine reaction rates and molecular opacities of sulfur-bearing species.

\end{abstract}

\section{Introduction}
\label{sec:intro}

\begin{figure*}
\centering
\includegraphics[width=0.95\textwidth]{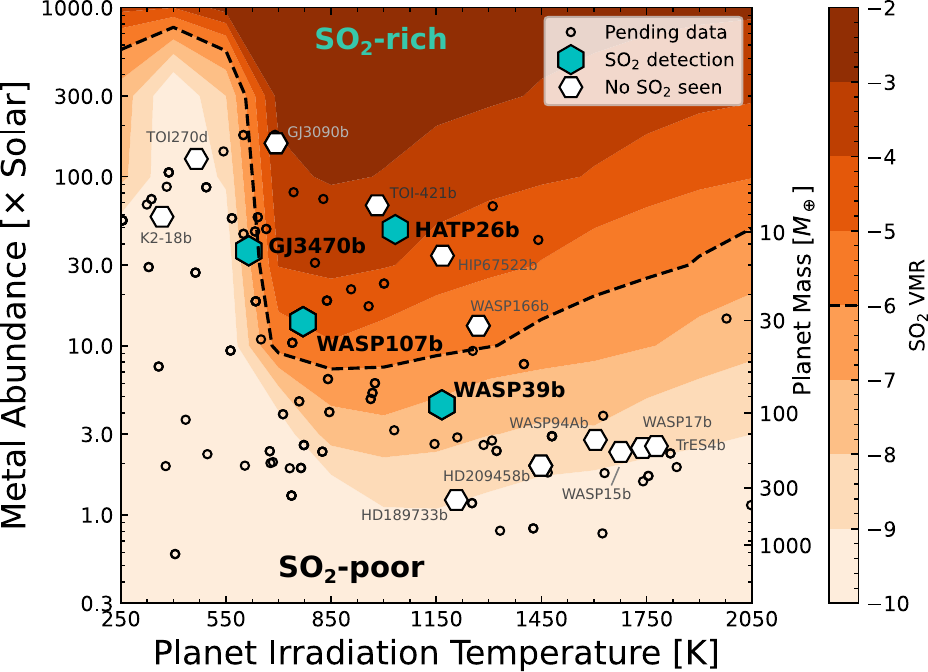}
\caption{\soo\ abundance averaged from 1--100\,$\mu$bar for our
  nominal model grid (color map), with the 1~ppm volume mixing ratio
  indicated by the dashed line.  We term this the ``\soo\ shoreline,''
  below which \soo\ is unlikely to be detectable.  Markers show the
  shoreline in the context of exoplanets observed by JWST in transit
  with $R_P\ge 1.8 R_\oplus$.  Empty symbols indicate unpublished
  observations; white hexagons show \soo\ nondetections while blue
  hexagons show \soo\ detections. {\referee All four non-detections within the \soo-rich region still allow for the presence of significant levels  of \soo.} }
\label{fig:shorelinemap}
\end{figure*}

\begin{deluxetable*}{l l l l l}[bt]
\tabletypesize{\scriptsize}
\tablecaption{  Model Parameters: \label{tab:params}}
\tablewidth{0pt}
\tablehead{
\colhead{Name} & \colhead{Units} & \colhead{Description} & \colhead{Value} & \colhead{Source} 
}
\startdata
\multicolumn{5}{l}{\hspace{0.1in}\em System parameters:}\\
     $R_*$ &  $R_\odot$ & Stellar radius & 0.79  &  \citep{wakeford:2017} \\
$T_\mathrm{eff}$ & K & Stellar effective temperature  & 5062 & \citep{wakeford:2017}     \\
     $R_P$ &  $R_\oplus$ & Planetary radius & 6.33 &  \citep{wakeford:2017} \\
     $M_P$ &  $M_\oplus$ & Planetary mass &  18.73&  \citep{wakeford:2017} \\
     $g_P$ &  m~s$^{-2}$ & Planetary surface gravity & 4.58  &  \citep{wakeford:2017} \\
\multicolumn{5}{l}{\hspace{0.1in}\em Modeling parameters: constant values}\\
He/H  & -- & Solar volume mixing ratio & $8.38 \times 10^{-2}$  & \cite{lodders:2020}\\
C/H  & -- & Solar volume mixing ratio & $2.95 \times 10^{-4}$  & \cite{lodders:2020}\\
N/H  & -- & Solar volume mixing ratio & $7.08 \times 10^{-5}$   & \cite{lodders:2020}\\
O/H  & -- & Solar volume mixing ratio & $5.37 \times 10^{-4}$   & \cite{lodders:2020}\\
S/H  & -- & Solar volume mixing ratio & $1.41 \times 10^{-5}$   & \cite{lodders:2020}\\
$P$ & bar & Pressure range & \multicolumn{2}{l}{1000--10$^{-7}$  } \\
$P_0$ & bar & Reference pressure & \multicolumn{2}{l}{0.01  } \\
$z$ & deg & Zenith angle & \multicolumn{2}{l}{83} \\
\multicolumn{5}{l}{\hspace{0.1in}\em Modeling parameters: varied values}\\
M/H  & -- & Metallicity relative to Solar & \multicolumn{2}{l}{0.3, 1, 3, 10, 30, 100, 30, 1000} \\
$T_\mathrm{eq}$ &    K &  Equilibrium temperature & \multicolumn{2}{l}{250, 400, 550, 700, 850, 1000, 1150,} \\
              &     &   & \multicolumn{2}{l}{\ 1300, 1450, 1600, 1750, 1900, 2050} \\
$a$ &    AU &  Semimajor axis & \multicolumn{2}{l}{0.753, 0.294, 0.156, 0.0960, 0.0651, 0.0471, 0.0356,} \\
 &  &   & \multicolumn{2}{l}{\ 0.0278, 0.0224, 0.0184, 0.0154, 0.0130, 0.0112} \\
$T_\mathrm{int}$ & K & Internal temperature & \multicolumn{2}{l}{{\bf 100}, 300, 500} \\
$F_\mathrm{XUV}$ & -- & Relative XUV irradiation & \multicolumn{2}{l}{0.03, {\bf 1.0}, 30} \\
$K_\mathrm{zz}$ & $\mathrm{~cm}^2\mathrm{\ s}^{-1}$ & Vertical diffusion coefficient & \multicolumn{2}{l}{$10^5$, {$\mathbf{10^7}$}, $10^9$} \\
 $C/O$ & -- & C/O ratio & \multicolumn{2}{l}{0.3, {\bf 0.55}, 0.8} \\
\enddata
\tablenotetext{}{Entries in {\bf bold} indicate the values used for the nominal model.}
\tablenotetext{}{All temperature profiles, chemical
profiles, and synthetic spectra are available electronically at \url{https://doi.org/10.5281/zenodo.17101615}}
\end{deluxetable*}

The unexpected presence of sulfur dioxide (\soo) has been one of the
biggest surprises to emerge from JWST's early spectroscopy of gas
giants.  This molecule burst onto the scene with the Early Release
Science transmission spectroscopy of WASP-39b
\citep{alderson:2023,rustamkulov:2023}.  \soo\ was soon found to
result from at least two separate disequilibrium photochemistry
pathways at high altitude in hydrogen-rich atmospheres, catalyzed by
high-energy ($\lambda < 230$\,nm)  stellar photons that convert
dissociate parent species to ultimately form
\soo\ \citep{tsai:2023,polman:2023,degruijter:2025,veillet:2025}.  This exciting
new absorber has since been seen in planets with sizes from
0.4--1.6\,$R_J$ and temperature from 600-1700\,K
\citep{beatty:2024,dyrek:2024}.  Its triatomic structure
makes it a sensitive metallicity tracer, and its photochemical origin
means it may also diagnose vertical mixing (e.g., via the eddy
diffusion coefficient $K_{zz}$), internal heat ($T_\mathrm{int}$), {\referee UV-induced}
photochemistry, atmospheric dynamics, and planet formation history
\citep{tsai:2023,crossfield:2023,khorshid:2024,constantinou:2023}.
However, many mysteries remain.

To date \soo\ has been detected in only {\referee four} exoplanets. Though first
observed at its 4.1\,\micron\ ($\nu_1+\nu_3$) combination band
\citep{alderson:2023,tsai:2023,rustamkulov:2023}, \soo\ and the
physical conditions it diagnoses also reveal themselves at the
longer-wavelength 7--9\,\micron\ $\nu_1$ and $\nu_3$ stretch vibration
bands \citep{powell:2024}.
While \soo\ is seen in hot Jupiters WASP-39b and WASP-107b, {\referee and in the smaller HAT-P-26b}, at both
4.1\,\micron\ and
7--9\,\micron\ \citep[][Alderson et al.\ in prep.]{powell:2024,alderson:2023,welbanks:2024,sing:2024,dyrek:2024,valenti:2025,gressier:2025},
the only other planet with detected \soo, warm Neptune GJ~3470b,
exhibits it in the weaker 4.1\,\micron\ band but lacks
longer-wavelength data \citep{beatty:2024}.

Beyond \soo, \hhs\ (with no associated \soo) has been seen in the
canonical hot Jupiter HD~189733b
\citep{fu:2024,inglis:2024,zhang:2025}. There have been a number of
other tentative hints of sulfur chemistry in exoplanet atmospheres
\citep[e.g.,][]{bannerjee:2024,benneke:2024,ahrer:2025,meech:2025,davenport:2025,felix:2025,mayo:2025}, but the number of confident detections of
such species remains low.


Precisely measuring an exoplanet's sulfur content allows us to finally
move beyond the hoary old C/O ratio, which has degenerate implications
for planetary formation conditions;
\citep[e.g.,][]{mordasini:2016,molliere:2022,feinstein:2025}. Unlike volatiles such
as C or O, sulfur has a much higher condensation/vaporization
temperature \citep{lodders:2003} and so it traces an exoplanet's
refractory budget.  S is particularly useful because it is the most
volatile of the major refractory constituents.  Rock and metal tracers
like Fe, Mg, and Si are only observable in the hottest of planets
\citep[$>$2000\,K;][]{lothringer:2021}, because they condense out of
the atmosphere for the vast majority of giant planets. Unfortunately,
these ultra-hot planets are rare and also suffer from complications
like molecular dissociation and ionization
\citep{arcangeli:2018,lothringer:2018b}, which make robust elemental
abundances challenging.  Thus, measuring sulfur in our targets
represents the best chance at tracing the accretion of refractory
material in giant planets \citep{crossfield:2023}.  Specifically, the
volatile-to-refractory C/S and O/S ratios may be the key to
distinguishing between pebble- or planetesimal-dominated accretion
\citep{turrini:2021,schneider:2021b,pacetti:2022} in a way that the oft-studied C/O
ratio cannot \citep{crossfield:2023}.

 Furthermore, sulfur as a refractory tracer informs us about other solids accreted onto the planet. For example, carbon fractionation in the disk: C in disk PAHs is released into the gas phase at the so-called soot line of protoplanetary disks \citep{kress:2010}. Based on  solar system meteorites, $\sim$50\%  of the disk carbon is found in the solid phase \citep{bergin:2015}, and so  ice phase carbon alone cannot provide an accurate estimate of the C budget in a disk. Ignoring the refractory carbon can result in inaccurate predictions for the planet formation based on its atomic composition.  For regions where the temperature reaches 1200\,K for a few hundred years, solid phase C will be released into the gas phase; this location is typically $\sim$0.3--3\,AU \citep{binkert:2023}.
 Simulations  predict  a direct connection between the amount of solid-phase C and the planet's {total} C and S abundances \citep[e.g.,][and references therein]{feinstein:2025}. Even though the chemical history and complex carbon chemistry can both cause uncertainties in the carbon fractionation inferred from sulfur abundances, {only observational data can provide the tightest possible C \& S constraints -- which  are essential to best reveal the history of planets and their natal disks.}

By observing large numbers of planets and enabling the identification
of SO$_2$ and \hhs\ in exoplanetary spectra (see Fig.~\ref{fig:shorelinemap}, JWST opens a new window
into the elemental makeup and disequilibrium processes that influence
giant planet compositions. In {\em equilibrium} chemistry SO$_2$ is
not the main bearer of atmospheric sulfur --- that is
\hhs\ \citep{polman:2023}. \soo's {\em photochemical} origin makes
\soo\ a distinct tracer of disequilibrium chemistry processes, being
sensitive to {\referee UV} flux, $K_{zz}$, $T_\mathrm{int}$, and global
circulation
\citep{tsai:2023,tsai:2023b,crossfield:2023,mukherjee:2025,degruijter:2025}.

SO$_2$ in giant planet atmospheres also traces enhanced atmospheric
metallicity \citep{tsai:2023,crossfield:2023}.  The \soo\ absorption
in the 7.4\,$\mu$m and 8.7\,$\mu$m bands are much more sensitive metallicity
tracers than those at 4.1\,\micron, which saturate at lower
metallicity \citep{tsai:2023,crossfield:2023}.  Thus longer-wavelength
 measurements of SO$_2$ are crucial to precisely measure the S
content (and for the total oxygen abundance, as \soo\ locks oxygen
away from CO, \coo\, and \water).

In this paper we calculate and present a series of atmospheric models
to better elucidate the location of the exoplanetary ``\soo
shoreline'' --- the boundary between regions of gas giant parameter
space where \soo\ is abundance enough to be detected, and regions
where it may go unseen. We also explore more generally the sulfur
chemistry of these H$_2$/He-dominated atmospheres.
Sec.~\ref{sec:modeling} presents our modeling
approach. Sec.~\ref{sec:discussion} discusses these results to
determine which parameters have a particularly strong or weak
influence on the expected atmospheric \soo\ abundance.
Sec.~\ref{sec:budget} then discusses the overall sulfur budget and the
dominant sulfur-carrying molecules expected in exoplanet atmospheres,
before concluding in Sec.~\ref{sec:conclusion}.


\begin{figure}
\centering
\includegraphics[width=0.5\textwidth]{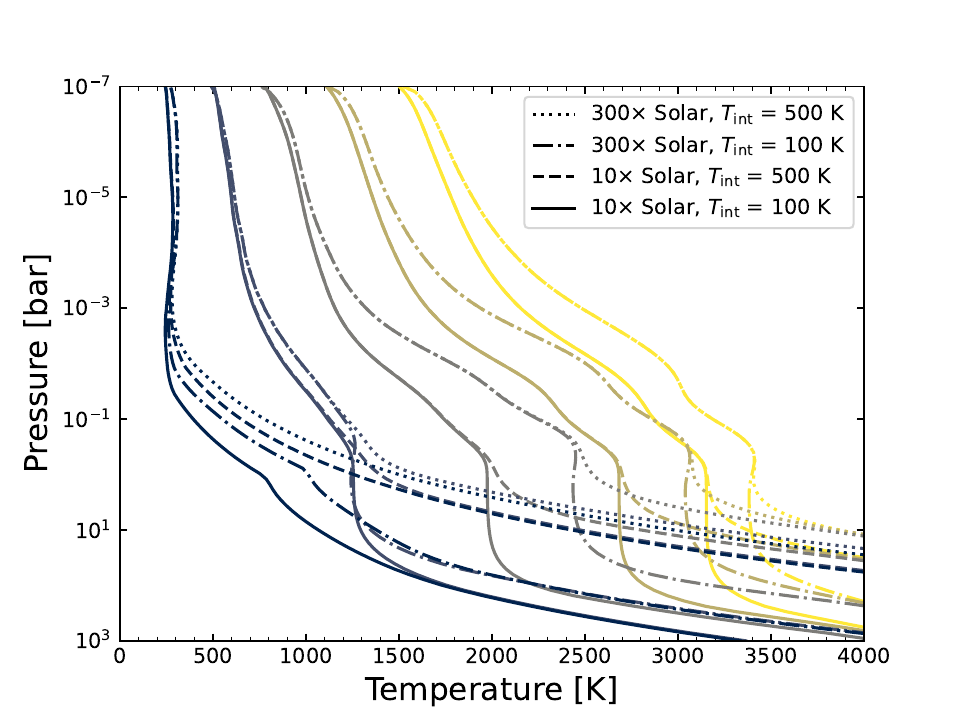}
\caption{Representative temperature profiles of our models, showing
  two metallicities and two $T_\mathrm{int}$ for equilibrium
  temperatures of 250, 700, 1150, 1600, and 2050~K. Changing
  $T_\mathrm{int}$ affects the temperature profile at
  $P\gtrsim1$~mbar, at altitudes below where transmission spectra
  probe and where \soo\ is produced.}
\label{fig:tps}
\end{figure}

\vspace{0.5in}





\begin{figure*}
\centering
\includegraphics[width=0.95\textwidth]{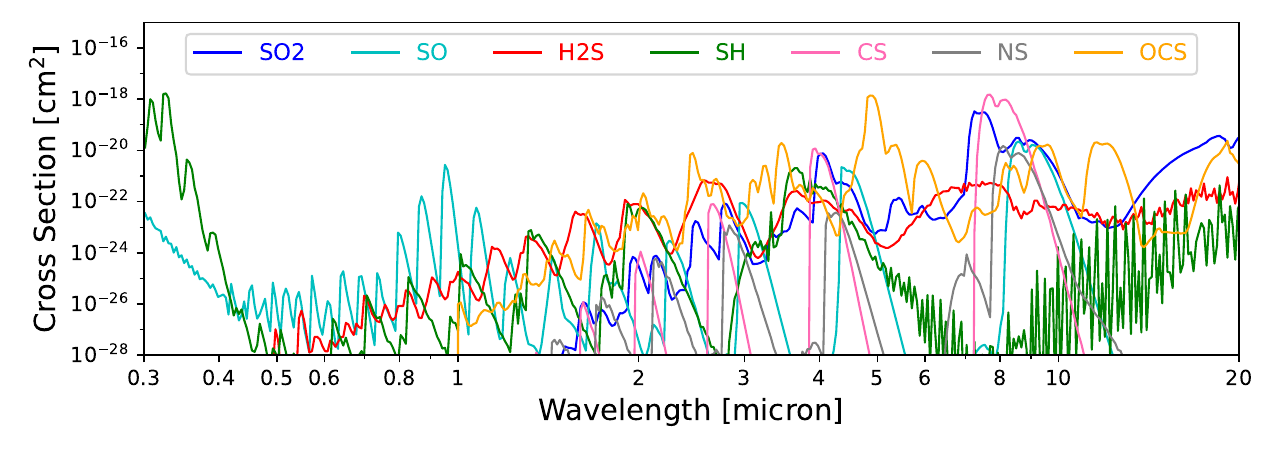}
\includegraphics[width=0.49\textwidth]{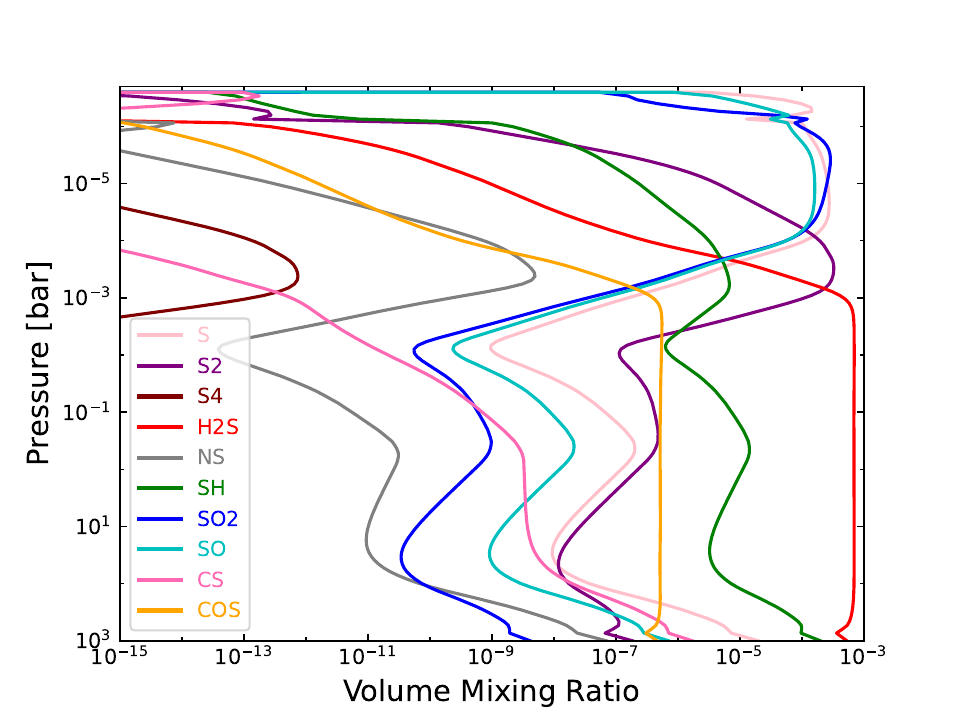}
\includegraphics[width=0.49\textwidth]{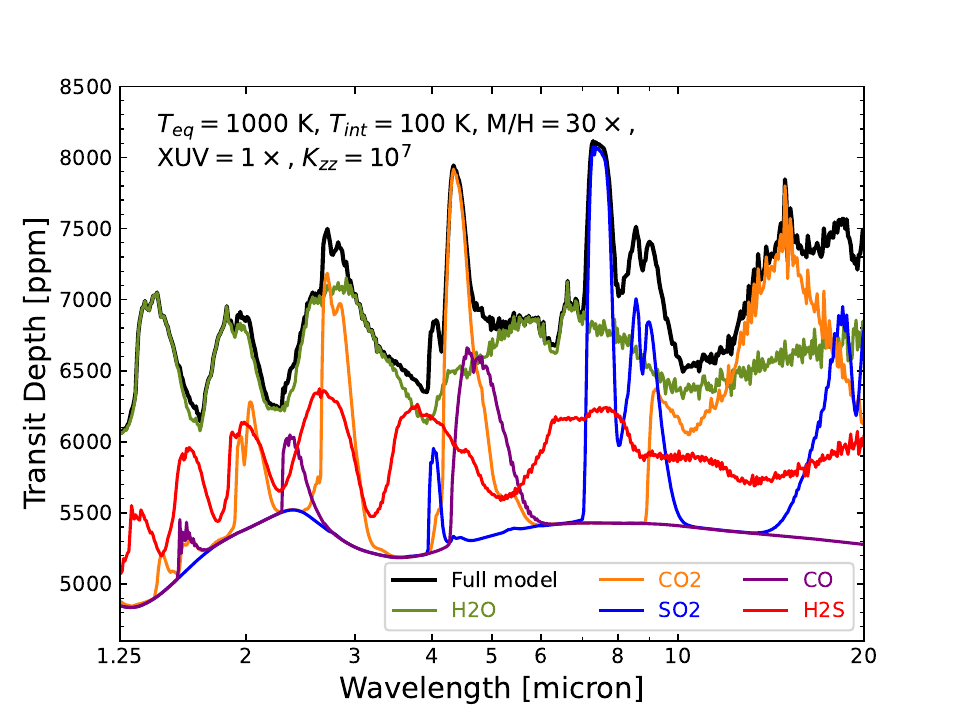}
\caption{{\em Top:} Cross-sections of sulfur-bearing species, calculated at 1000~K and 10~mbar. The data have been smoothed for clarity. {\em Bottom left:} Abundance profiles for \teq=1000\,K, $T_\mathrm{int}$=100 K, 30$\times$ Solar metallicity, and nominal {\referee high-energy} irradiation and C/O ratio.
{\em Bottom right:} Full synthetic transmission spectrum
   (black line) together with model spectra including only one molecular
   absorber at a time (colored lines).  Molecules not listed
   contribute negligibly to this  spectrum.}
\label{fig:crosssec}
\end{figure*}

\section{Modeling}
\label{sec:modeling}
To map the exoplanetary \soo\ shoreline, we generated a suite of model
atmospheres and corresponding synthetic spectra largely following the
same approach as \cite{crossfield:2023}.  Our modeling parameters are
summarized in Table~\ref{tab:params}.  In brief: we use
\texttt{HELIOS} \citep{malik:2017} to generate vertical temperature
profiles in radiative-convective {\referee (but not thermochemical)}
equilibrium for a given bulk metallicity (M/H), semimajor axis ($a$,
or equivalently equilibrium temperature $T_\mathrm{eq}$), internal
temperature ($T_\mathrm{int}$), and related system parameters
{\referee (\citeauthor{mukherjee:2025} \citeyear{mukherjee:2025}
  showed that fully self-consistent equilibrium profiles change
  temperatures by $<$10\%)}. Fig.~\ref{fig:tps} shows a subset of the
resulting temperature profiles; {\referee for all models we used a
  single heat-recirculation efficiency of 0.5, on a scale from 0.25 to
  1.0}. We then use \texttt{VULCAN} \citep{tsai:2017} with its SNCHO
chemical network to calculate vertical chemical profiles for each
\texttt{HELIOS} temperature profile for a range of XUV irradiation,
C/O ratio, and vertical diffusion ($K_\mathrm{zz}$).  Finally, we use
\texttt{petitRADTRANS} \citep{molliere:2019} to calculate synthetic
transmission spectra of the planets; the spectra include molecular
opacity from H$_2$O, CO, CO$_2$, OH, SO$_2$, SO, CH$_4$, HCN, H$_2$S,
NH$_3$, CH$_3$, C$_2$H$_2$, C$_2$H$_4$, CN, CH, SH, CS, H$_2$CO,
H$_2$O$_2$, N$_2$O, NH, NO, and NS as well as atomic Na and K at each
specified metallicity level.
Fig.~\ref{fig:crosssec} shows the typical cross-sections of some of
the key sulfur-bearing species, vertical mixing ratios for
sulfur-bearing species, and a representative model transmission
spectrum. References for our opacities are listed in the Appendix.

{\referee Since Na and K are not treated in \texttt{VULCAN}'s chemical
  network, to calculate our transmission spectra we simply scaled the
  abundances of these elements relative to the solar values in the
  same way as for C, N, O, and S.  Any species including other
  elements beyond those listed above are not included in the
  atmospheric chemistry or resulting spectra. Note that our analysis also
  does not account for phase changes; particularly for our coolest
  models, \water\ condensation may result in significant changes to
  the atmospheric thermal structure and chemistry compared to the
  results we present here.}

Since \soo\ has been detected in a wide range of planets, from the
smallish warm Neptunes GJ~3470b to {\referee HAT-P-26b} to the hot Jupiters WASP-107b and
WASP-39b \citep[][Alderson et al.\ in prep., Gressier et al.\ in prep.]{beatty:2024,alderson:2023,sing:2024,welbanks:2024,valenti:2025}, in
our models we adopt the properties of the intermediate-size planet
HAT-P-26b.  The physical system parameters we take from
\cite{wakeford:2017}.  Our nominal stellar spectrum for the
\texttt{VULCAN} photochemistry calculations is from the MUSCLES
Extension program \citep{france:2016,behr:2023}, which includes
observed STIS spectra of HAT-P-26, with X-ray and FUV data taken from
HD~40307 \citep{youngblood:2016}. Our models with semimajor axis
0.0471~AU (1000~K) correspond most closely to the actual planet
HAT-P-26b, while the remaining models span the rest of irradiation
parameter space.

The final result is several grids of model atmospheres and spectra.
Our nominal grid spans 13 semimajor axes (corresponding to
$T_\mathrm{eq}$ from 250~K to 2050~K) and 8 metallicities (from
0.3$\times$ to 1000$\times$ the Solar level) for an initial total of
104 models. We also calculate numerous grids in which a single
parameter is changed: $T_\mathrm{int}$ of 300~K or 500~K (vs.\ a
nominal value of 100~K); XUV irradiation at either 0.03$\times$ or
30$\times$ the nominal value; {\referee constant-with-altitude}
$K_\mathrm{zz}$ of $10^5$ or $10^9$ (nominal value $10^7$), or an
elemental C/O ratio of 0.30 or 0.80 instead of the nominal, Solar
value of 0.55. Table~\ref{tab:params} lists all these parameters, with
the nominal parameters in bold. {\referee Again, note that the lack of
  condensation treatment in our models may significantly impact the
  accuracy of the coolest set of models.} All temperature profiles,
chemical profiles, and synthetic spectra are available electronically
online\footnote{Zenodo link, to be uploaded and linked upon
publication.}.

\section{Results: Tracing the \soo\ Shoreline}
\label{sec:discussion}
Figure~\ref{fig:shorelinemap} shows the exoplanetary ``\soo\ shoreline,''
regions of temperature and metallicity where \soo\ is predicted to be
abundant. The color shading indicates the predicted amount of
\soo\ (averaged from 1--100~$\mu$bar) in the atmospheres of these
planets assuming \tint=100\,K, \kzz=$10^7$~cm$^2$~s$^{-1}$, and
C/O=0.55. The dashed line at a volume mixing ratio of $10^{-6}$
 roughly indicates the minimum abundance
 that we might hope to detect \citep[cf.][]{powell:2024,sing:2024}.

The \soo\ shoreline is a steep function of planet temperature and
metallicity \citep{polman:2023,tsai:2023,crossfield:2023}. Below
$\sim$600~K the \soo\ abundance is extremely low for M/H $\lesssim$
300$\times$ Solar, but for higher irradiation levels \soo\ is
considerably more abundant and may be detectable at metallicities as
low as $\sim$3$\times$ Solar. We next consider how the shoreline
advances or retreats as various parameters are adjusted.
\vspace{0.1in}

\vspace{0.1in}
\noindent{\bf Internal Temperature} Atmospheric \soo\ is particularly
insensitive to $T_\mathrm{int}$.  Fig.~\ref{fig:tps} shows that
increasing $T_\mathrm{int}$ from 100~K to 500~K only marginally
changes the temperature profile in the upper atmosphere where \soo\ is
produced.  Thus, we find that the \soo\ shoreline in our nominal model
($T_\mathrm{int}$=100~K) is almost indistinguishable from the
shoreline predicted when $T_\mathrm{int}$=500~K. This result
indicates that $T_\mathrm{int}$ is unlikely to be a confounding factor
when interpreting measured \soo\ abundances, confirming the results of
\cite{mukherjee:2025}.

\begin{figure*}
\centering
\includegraphics[width=0.475\textwidth]{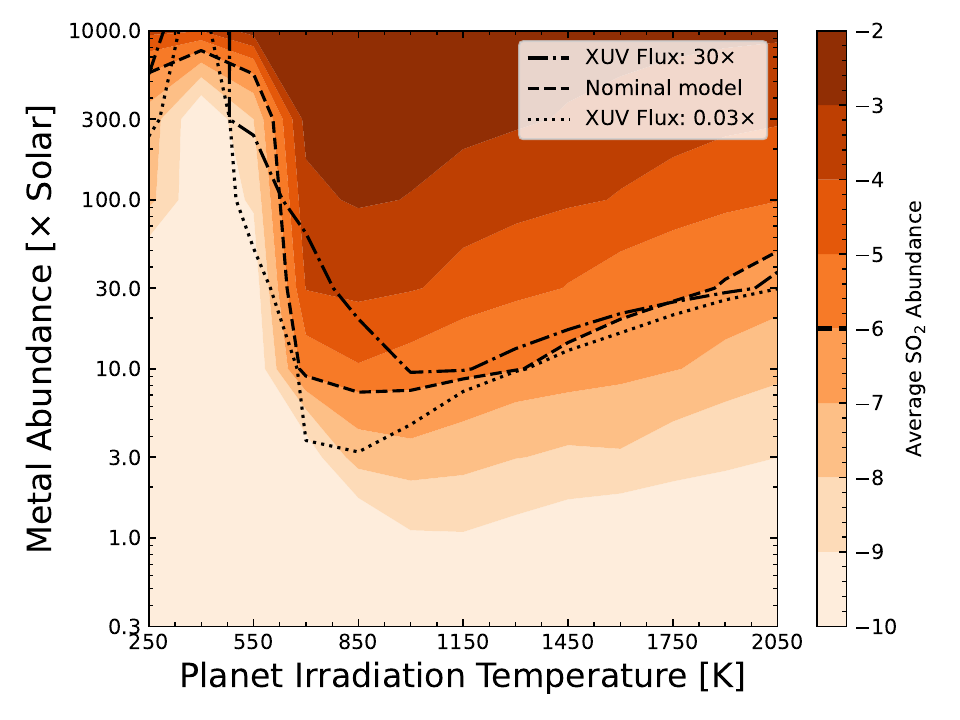}
\includegraphics[width=0.515\textwidth]{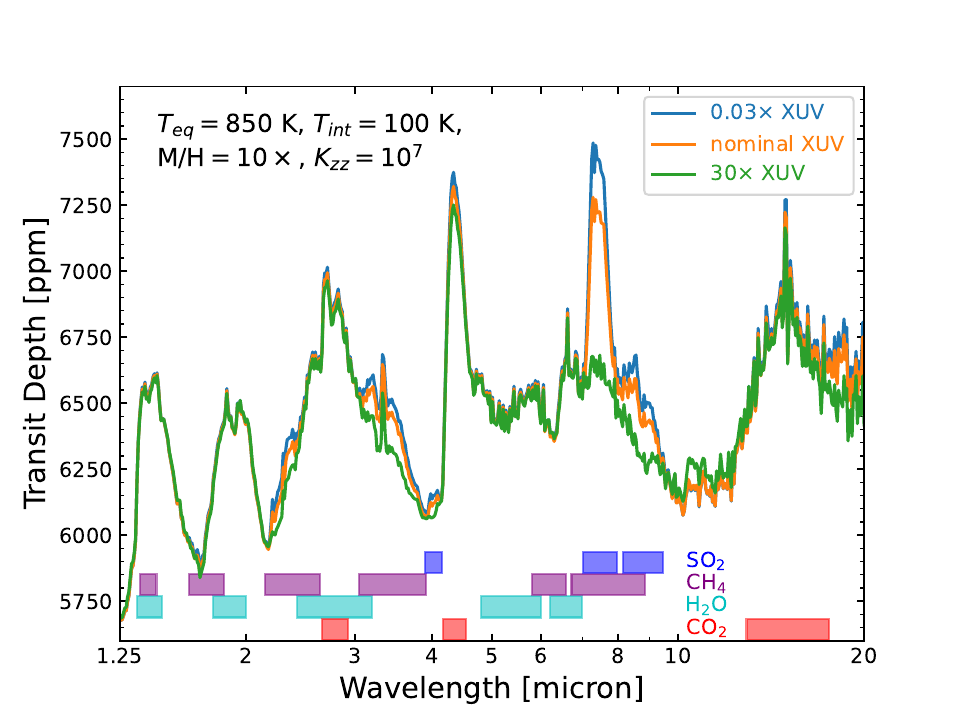}
\caption{{\em Left:} Same as Fig.~\ref{fig:shorelinemap}, but here showing the impact of varying the incident XUV irradiation.  Significantly higher or lower levels of XUV flux can  somewhat shift the \soo\ shoreline for temperatures $\lesssim$1400~K. {\em Right:} representative spectra showing that dramatically increasing the level of incident XUV results in lower levels of \soo\ (as well as \methane).}
\label{fig:so2_xuv}
\end{figure*}

\vspace{0.1in}
\noindent {\bf Incident XUV Flux:} Surprisingly, we also find a
relatively weak dependence of \soo\ abundance on the incident {\referee high-energy}
flux.  Fig.~\ref{fig:so2_xuv} reveals that changing the $\lambda
< 230$\,nm flux by a factor of 30 negligibly shifts the
\soo\ shoreline for \teq$\gtrsim$1000\,K, with somewhat larger effects
at cooler temperatures.  At a given temperature and metallicity,
significantly increasing the {\referee high-energy} flux generally decreases the amount
of atmospheric \soo.  The result is surprising considering the
stronger dependence of \soo\ on incident XUV flux reported by previous
studies \citep{dyrek:2024,degruijter:2025}, {\referee but is consistent with that found by \cite{mukherjee:2025}.}
Since high-energy stellar flux is largely or entirely unmeasured for most systems
where \soo\, \hhs, and other sulfur-bearing species are sought, this
result implies that this uncertainty in stellar flux may have  at
most a modest impact on interpreting these \soo\ detections.

\begin{figure*}
\centering
\includegraphics[width=0.475\textwidth]{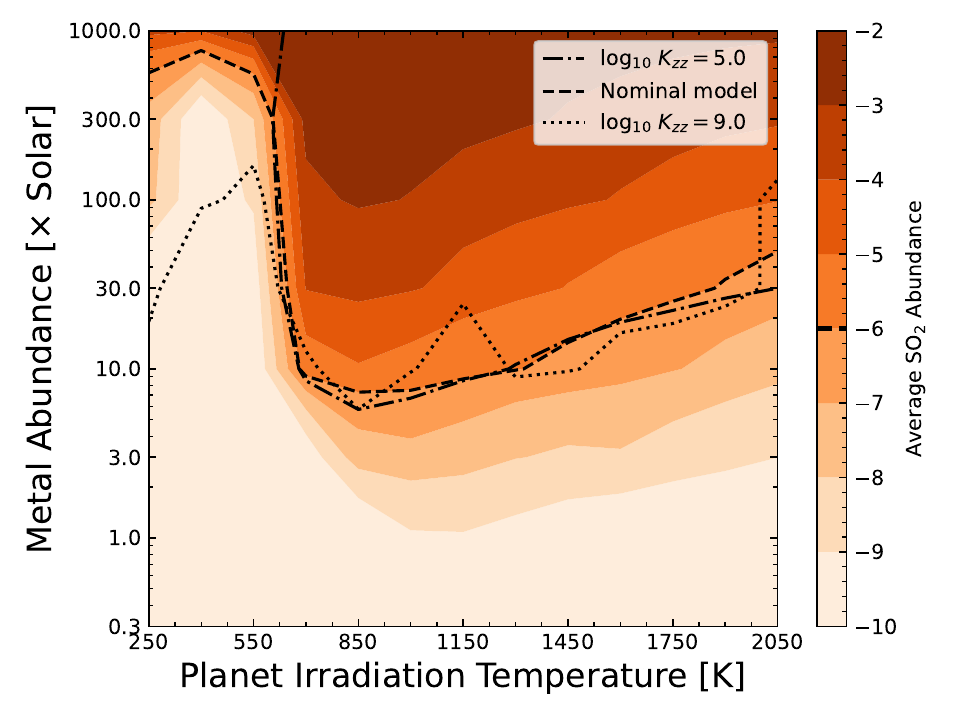}
\includegraphics[width=0.515\textwidth]{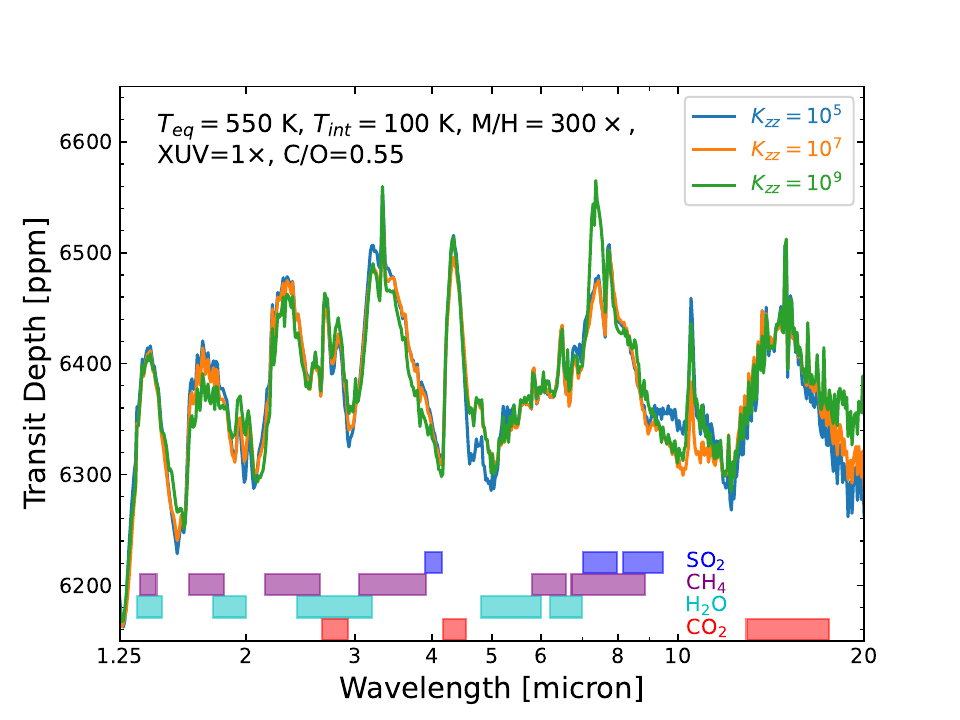}
\caption{Same as Fig.~\ref{fig:so2_xuv}, but here showing
  the impact of varying the vertical diffusion coefficient,
  $K_\mathrm{zz}$.  Varying \kzz\ over four orders of magnitude hardly
  budges the \soo\ shoreline for temperatures $\gtrsim$600~K; at lower
  temperatures, sufficiently high \kzz\ may result in measurable
  \soo\ for sufficiently high metallicities.
Representative spectra showing that increasing \kzz\ over this range results in a slightly more detectable  \soo\ signature at 7.4\,\micron.  {\referee Note also the prominent ethylene (C$_2$H$_4$) feature at 10.5\,\micron.}}
\label{fig:so2_kzz}
\end{figure*}

\vspace{0.1in}
\noindent {\bf Vertical Mixing:} The impact of $K_\mathrm{zz}$ is more
striking. Fig.~\ref{fig:so2_kzz} reveals that varying $K_\mathrm{zz}$
by four orders of magnitude results in negligible changes to the
\soo\ shoreline for $T_\mathrm{eq}\gtrsim 600$~K.  However, at cooler
temperatures a stark change emerges: {\referee below 600~K, a lower
  $K_\mathrm{zz}$ dramatically decreases the \soo\ abundance in the
  upper atmosphere at a given temperature and metallicity.}  This
effect is most likely to be detectable for high-metallicity
($\gtrsim100\times$ Solar) planets with $K_\mathrm{zz}\gtrsim10^7
\mathrm{~cm}^2\mathrm{\ s}^{-1}$ {\referee and $T_\mathrm{eq}
  \lesssim600$~K.  Fig.~\ref{fig:so2_kzz} also hints that for
  $T_\mathrm{eq} \gtrsim 1200$~K the trend may reverse, with higher
  \kzz\ implying lower \soo\ abundance; however, this effect seems
  weak at best and so it may be more difficult to observationally
  confirm.}

{\referee 
   At equilibrium temperatures $\lesssim 600$~K SO2 is not favored to
   form in significant amounts.  The cause of the \kzz\ dependence at
   those low temperatures seems to involve sufficiently strong
   vertical mixing that dredges \soo\ up from deeper regions of the
   atmosphere (below 1 bar) where it is more plentiful than it is in
   the atmospheric regions probed by transmission spectroscopy. There
   may also be an aspect that some of the photochemically-produced
   \soo\ is mixed downward to layers where it is less easily destroyed
   by processes in the thin upper atmosphere. The details of this
   dependence on vertical mixing will be the subject of a future
   study.}


\begin{figure*}
\centering
\includegraphics[width=0.475\textwidth]{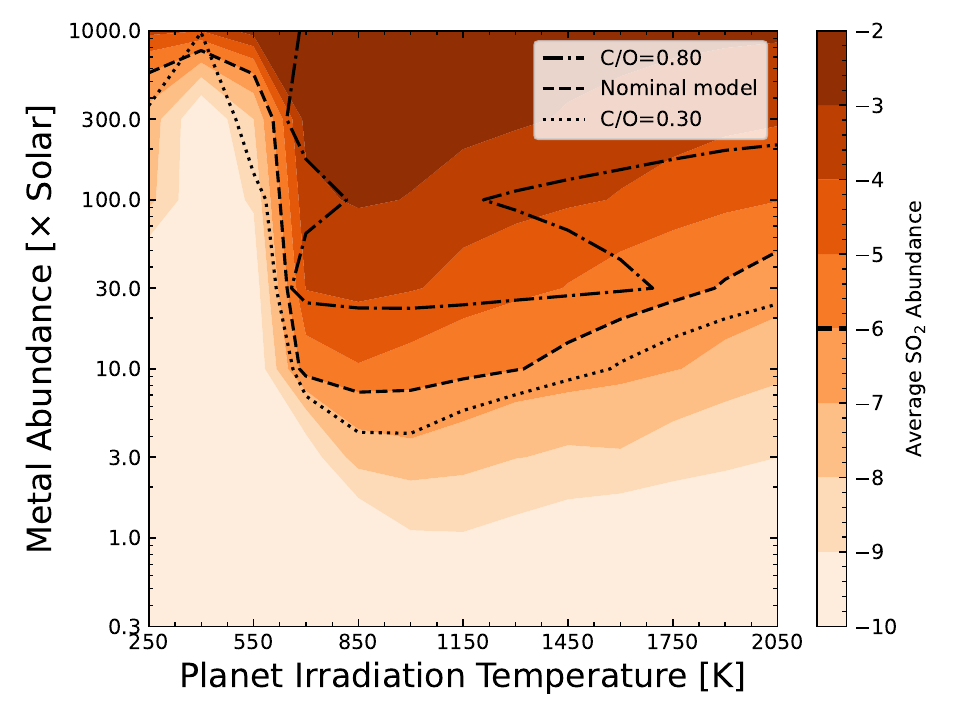}
\includegraphics[width=0.515\textwidth]{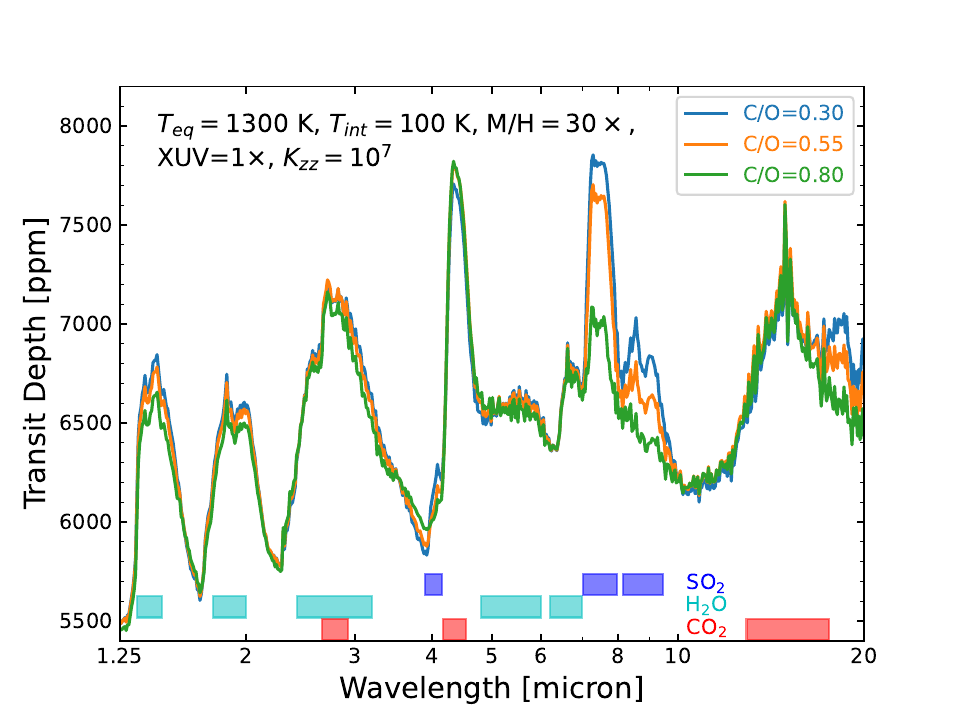}
\caption{{\em Left:} Same as Fig.~\ref{fig:shorelinemap}, but here showing
  the impact of varying the atmospheric C/O ratio. Increasing or
  decreasing the C/O ratio can significantly shift the \soo\ shoreline
  up or down, respectively.  {\em Right:} Representative spectra
  showing how decreasing the C/O ratio can significantly increase the
  atmospheric \soo signature, especially at 7.4\,\micron.}
\label{fig:so2_c2o}
\end{figure*}

\vspace{0.1in}
\noindent {\bf C/O Ratio:} Finally, Fig.~\ref{fig:so2_c2o} shows that
the often-sought atmospheric C/O ratio has the strongest effect on the
\soo\ shoreline and overall \soo\ abundance.  As the C/O ratio is
increased from 0.30 to 0.80, the \soo\ abundance steadily decreases
\citep[as seen in previous studies,
  e.g.][]{polman:2023,crossfield:2023,beatty:2024,mukherjee:2025}. Notably, the
C/O ratio is the only effect we explored that shifts the
\soo\ shoreline by roughly equal amounts at all planet temperatures.  {\referee We also note that the shape of the shoreline appears to qualitatively change at the higher C/O ratio of 0.8; future studies may be in order to determine the cause of this change.}

{\referee Regardless, the} detection of \soo\ is therefore a strong
sign of either a C/O ratio $\lesssim$ the Solar value, and/or a high
overall metallicity. If most giant planets have C/O comparable to
their host stars, then their C/O will be on average slightly
supersolar {\referee \citep[because the average stellar C/O is
    supersolar;][]{fortney:2012}} and the planets' expected
\soo\ abundances will be somewhat decreased relative to our nominal
conditions {\referee (because higher C/O tends to imply lower
  \soo\ abundances; cf.\ Fig.~\ref{fig:so2_c2o}).}

Finally, we note that C/O is not the only elemental ratio relevant for
predicting atmospheric sulfur abundances. Ratios such as C/S or O/S
\citep[perhaps proxies for the volatile-to-refractory
  ratios;][]{lothringer:2021} are also relevant for determining the
atmospheric abundances of sulfur-bearing species
\citep{khorshid:2024,crossfield:2023}.



\begin{figure*}
\centering
\includegraphics[width=0.9\textwidth]{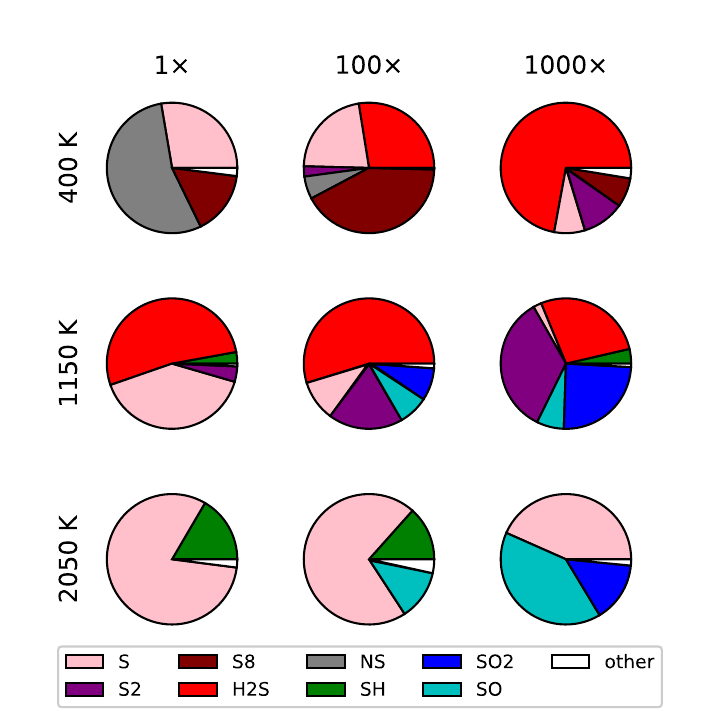}
\caption{Each chart shows the fractional contribution of the indicated species to the overall atmospheric sulfur budget at the noted temperature and metallicity.  The calculation indicates an average from 10$^{-5}$ to 10$^{-2}$~bar. }
\label{fig:pies}
\end{figure*}

\section{Where has all the sulfur gone?}
\label{sec:budget}

Fig.~\ref{fig:pies} shows the overall atmospheric sulfur budget of
H$_2$/He-dominated atmospheres across a wide range of temperatures and
metallicities, as predicted by our atmospheric models. Despite the
considerable attention recently directed at \soo, in only a small
subset of cases is \soo\ one of the most abundant sulfur-bearing
molecules.  H$_2$S, S$_2$, NS, SO, SH, and even S$_8$ are frequently
as common (or more so) as \soo\ depending on the overall temperature
and heavy element enrichment level. Contrary to \cite{mukherjee:2025},
we do not find that CS and CS$_2$ ever carry a substantial fraction of
the atmospheric S.  {\referee Note however that this is likely because
  \texttt{VULCAN}'s chemical network currently lacks some reactions
  involving a subset of C-S bonded species
  \citep[cf.][]{moses:2024,veillet:2025}. However, we do find that at
  lower metallicities and intermediate equilibrium temperatures
  (400--600 K) the sulfur that is in SO2 at higher temperatures
  becomes sequestered in CS$_2$, OCS, and S$_2$.  At the lowest
  temperatures in our model grid, at low-to-moderate metallicities the
  sulfur is increasingly in higher-order allotropes such as S8.}


Fig.~\ref{fig:crosssec} demonstrates the
relative detectability of many of these sulfur-bearing species.
Consistent with previous work \citep[e.g.,][]{polman:2023,tsai:2023},
we find \soo\ to be the most readily discerned such molecule: its
opacity is greatest at 7.4\,\micron\ band (and, to a lesser extent, at
8.7\,\micron), but observational S/N is typically greater at the
4.1\,\micron\ band. \hhs\ is often more abundant than \soo\ but strong
detection of \hhs\ is much more challenging than for \soo\ in
transmission; the best prospect may be to catch a glimpse of its broad
3.7\,\micron\ band in between the typically-stronger bands of
\water\ and/or \methane\ below, and \soo\ and/or \coo\ above --- as
done for HD~189733b in transmission \citep{fu:2024}. Unfortunately,
half of this \hhs\ signature lies in an unobserved detector gap when
using JWST's NIRSpec/G395H mode, both partially obscuring the signal
and potentially inducing systematic transit-depth offsets between the
two NIRSpec detectors \citep[cf.\ ][]{ahrer:2025b}. Alternatively,
\hhs\ is also detectable in emission spectra from
7--8\,\micron\ \citep{inglis:2024}. Finally, SO achieves appreciable
abundances at higher metallicities and temperatures, and may be
visible from 0.8--1.2\,\micron.











\section{Conclusion}
\label{sec:conclusion}

\subsection{Summary of Results}
Sulfur chemistry remains elusive in exoplanet atmospheres. To date
\soo\ has been detected in only four exoplanets. Though first observed
at its 4.1\,\micron\ ($\nu_1+\nu_3$) combination band
\citep{alderson:2023,tsai:2023,rustamkulov:2023}, \soo\ and the
physical conditions it diagnoses also reveal themselves at the
longer-wavelength 7--9\,\micron\ $\nu_1$ and $\nu_3$ stretch vibration
bands \citep{powell:2024}. \hhs\ has been seen confidently only via emission spectroscopy, also at 7--8\,\micron\ \citep{fu:2024,inglis:2024}.

With only a handful of planets having detections of sulfur-bearing
species, more observatoinal data are urgently needed. JWST will remain
the superlative facility for studying \soo\ in exoplanet atmospheres.
ESA's Ariel (scheduled for launch in 2029) will also observe \soo\ in
transiting exoplanet atmospheres via AIRS Channel 1, which will
provide low-resolution spectra from 3.9--7.8\,\micron. These data
should easily measure \soo\ at 4.1\,\micron, and the
longest-wavelength data ($\gtrsim$7\,\micron) may also be sufficient
to see \soo. These facilities, and perhaps also ground-based
telescopes, could also descry SO and/or SH at wavelengths below
$\lesssim$1.2\,\micron.

Our modeling maps the expected location of the \soo\ shoreline, the
boundary between atmospheres with or without detectable \soo.  Under
our nominal model assumptions (Table~\ref{tab:params}) \soo\ should be
detectable for $T_\mathrm{eq}$$\gtrsim$600~K and metallicities
$\gtrsim$10$\times$ Solar. the location of the shoreline is
insensitive to \tint, depends weakly on \kzz\ and XUV irradiation, and
varies particularly strongly when the C/O ratio is
adjusted. Fig.~\ref{fig:shorelinemap} compares our nominal
\soo\ shoreline to the properties of transiting planets being observed
by JWST and finds decent agreement; however, data on more
intermediate-metallicity planets (and higher-S/N data on the smallest,
highest-metallicity planets) is needed to fully validate our
predictions.

\subsection{Discussion}
Indeed, further observations of a large number of planets are the
best hope for determining the true boundary of our proposed \soo\ shoreline.
Fig.~\ref{fig:shorelinemap} shows the \soo\ shoreline in the context
of all planets being observed by JWST in transit through Cycle~4.  In
this figure planet mass is mapped to metallicity assuming the C/H
mass-metallicity relation of \cite{welbanks:2019}.

Of the four planets with clear \soo\ detections, three --- {\referee HAT-P-26b,}
WASP-107b, and WASP-39b --- sit within or near to the nominal
\soo\ shoreline
\citep{beatty:2024,tsai:2023,powell:2024,welbanks:2024,sing:2024}. While
the previously-studied WASP-39b and WASP-107b sit comfortably within or near
the shoreline, GJ\,3470b sits right on the edge.
The possible discrepancy could be resolved if GJ~3470b's atmospheric
C/O is substantially non-solar and/or if it does not follow the
putative mass-metallicity trend assumed in
Fig.~\ref{fig:shorelinemap}.

The lower-mass planets with non-detections of \soo\ are perhaps more
intriguing (Fig.~\ref{fig:shorelinemap}).  All {\referee four} lie above the
\soo\ shoreline, and all show tentative hints of \soo\ just short of a
significant detection. These include the 32\,$M_\oplus$ WASP-166b
(VMR$\sim$10~ppm) \citep{mayo:2025}, {\referee }the 14\,$M_\oplus$ HIP 67522b \citep[VMR $\lesssim$1\,ppm][]{thao:2024}, the 7.2\,$M_\oplus$ TOI-421b
(1--10\,ppm) \citep{davenport:2025}, and the low-mass
(3.3\,$M_\oplus$) GJ~3090b \citep[1-100\,ppm;][]{ahrer:2025}. In all
cases the precision of the reported \soo\ abundances are broad enough
that the atmospheric compositions may still be generally consistent
with our model predictions, but higher signal-to-noise data are needed
to test our models.

Finally, the large number of massive hot Jupiters
\citep[e.g.,][]{kirk:2024,ahrer:2025} lacking definitive
\soo\ detections seems consistent with the model outlined in
Fig.~\ref{fig:shorelinemap}. If this population of large, massive
planets is consistently low-metallicity then {\referee we may expect little or no detectable \soo\ in many of the largest, highest-S/N exoplanetary targets.}

Besides comparing to individual planets, our models here will also
support research efforts on other exoplanetary fronts.  Our grid
provides 72 model atmospheres and corresponding spectra at
\teq=1000\,K, appropriate for interpreting upcoming JWST spectroscopy
of the wide range of planets marked in Fig.~\ref{fig:shorelinemap}.
Our total grid includes 936 models, which may be a sufficiently large
set to enable testing of machine-learning approaches on exoplanetary
chemistry and spectra across a wide range of parameter space.

\subsection{Future Work and Caveats}
Beyond  these results, our work leaves room for
improvement on several fronts.

First, our models include no clouds or hazes; such materials can
significantly impact a planet's resulting transmission spectrum, as
well as sequestering elements out of the gas phase into solids.  The
most likely sulfur bearing condensate clouds are ZnS, MnS, and Na$_2$S
(although there are also good reasons to think that the latter
  two do not form). The formation of such condensates is
limited by the abundance of the trace metals Zn, Mn, and Na; the
process will typically deplete $\lesssim$10\% of the bulk sulfur, but
not typically enough to significantly alter the atmospheric
\soo\ abundance \citep{tsai:2023,mukherjee:2025}. Sulfur polymerized into the S$_8$
molecule can form hazes for \teq$\lesssim$700~K, at least at
near-Solar metallicities \citep{gao:2017}; though by itself not enough
to appreciably change a planet's atmospheric sulfur budget,
condensation of this allotrope might even result in significant
reduction in sulfur-species abundances above the haze layer.

{\referee Phase changes of \water\ were also not inclued in our
  current analysis, even though Fig.~\ref{fig:tps} shows that for our
  coolest models, atmospheric temperatures drop below the condensation
  temperature of \water.  Such phenomena are likely to have a
  significant impact on both the thermal structure and chemistry of
  these atmospheres, and so this set of models and their associated
  chemistry and spectra may differ substantially from the results of a
  more complete and self-consistent analysis.}

In a related direction, our use of \texttt{VULCAN}'s SNCHO chemical
network restricts the applicability of our results in regions of
parameter space where reactions with other elements are
important. Thus the silicate-bearing clouds inferred for a growing
number of hot Jupiters \citep{dyrek:2024,inglis:2024} remain
unacounted for in our analysis.  Recent claims of silicon-bearing
gas-phase chemistry in H$_2$-dominated atmospheres \citep{ma:2025} are
untested with our current framework, as are the effects of TiO and
other molecules bearing heavier atoms
\citep[cf.][]{prinoth:2022}. {\referee Furthermore, \texttt{VULCAN}'s
  chemical network currently lacks some C-S bonded species, which may
  result in an overestimate of \soo\ at cooler temperatures
  where C-S species could take up a larger fraction of the sulfur
  \citep{veillet:2025,mukherjee:2025}.  The details of the SO$_2$ shoreline may
  change somewhat when considering those relevant reactions.}

Furthermore, our approach is not fully self-consistent inasmuch as
each model's chemistry is computed assuming a fixed, input thermal
profile.  In reality the thermal profile and atmospheric composition
necessarily evolve together: when one changes, the other changes as
well.  As \cite{mukherjee:2025} point out, the effects of this choice
are not fatal when considering transmission spectra and overall trends
in exoplanetary atmospheric abundances.

Our treatment of atmospheric chemistry with one-dimensional models is
overly simplistic since planets are, of course,
three-dimensional. Models linking atmospheric dynamics to chemistry
indicate that atmospheric abundances can change significantly between
the morning and evening terminator, to say nothing of differences seen
at other longitudes \citep{lee:2023,tsai:2024}. Similar studies could
also test current models by simulating the time-dependent atmospheric
chemistry expected in planets on eccentric orbits
\citep[e.g.,][]{lewis:2013,tsai:2023c}.

Analyses such as ours should also be extended to a wider
range of planet surface gravities than the single, HAT-P-26b-like,
case explored here \citep[or the WASP-39b-like gravity used in the
  model grids of ][]{crossfield:2023,mukherjee:2025}.  Lower surface
gravity can itself result in the formation of additional \soo, since in
such atmospheres the physically thicker atmosphere can provide additional
shielding against \soo\ photodissociation \citep{degruijter:2025}.

Uncertainties also remain as to the exact photochemical pathways
setting the observable sulfur species.  \cite{degruijter:2025} only
recently identified a new pathway for \soo\ formation beyond that
initially reported \citep{polman:2023,tsai:2023}.  Additional
explorations in this direction, whether for \soo\ or other species,
could be fruitful.

There is also the need for considerably more labaroratory and
theoretical calculations, since our model spectra still lack several
key opacity sources and/or reaction rates. Opacity data exist for
H$_2$CS, but it is not present in \texttt{VULCAN}'s SNCHO chemical
network. Many more molecules are in the network but lack opacity data
in public databases such as ExoMol or DACE: S$_2$, S$_3$ , S$_4$ ,
S$_8$, HS$_2$, HCS, HSO, HSO$_3$, H$_2$SO$_4$, CH$_3$S, CH$_3$SH,
S$_2$O.  And other sulfur-bearing molecules, such as sulfanes and
hydropolysulfide (H$_2$S$_n$) may also be relevant \cite{zahnle:2009}
in some regions of planetary parameter space. Obtaining such
reaction-rate and opacity data in a format compatible with public
modeling tools will be essential to determine the relative importance
and detectability of all sulfur-bearing species.



\vspace{-0.1in} \acknowledgments

We thank the anonymous referee, as well as Dr.\ J.\ Moses, for useful
comments and suggestions that significantly improved the final manuscript.  We also thank Drs.\ L.\ Alderson, Th.\ Beatty, and J.\ Valenti for useful discussions at the 2025 ExoClimes conference.


J.P. acknowledges the financial support of the SNSF under grant 51NF40\_205606.

C.P.-G. acknowledges support from the E. Margaret Burbidge Prize Postdoctoral Fellowship from the Brinson Foundation.

{\em Facilities:} JWST, ARIEL

{\em Software:} \texttt{VULCAN} \citep{tsai:2017}, \texttt{petitRADTRANS} \citep{molliere:2019}, \texttt{HELIOS} \citep{malik:2017}

\appendix

Table~\ref{tab:opacity} lists the sources used for our molecular
opacities. Most of these were taken from the ExoMol project
\citep{chubb:2020}, with a smaller number from MoLLIST
\citep{bernath:2020} and one (NO) from HITEMP \citep{rothman:2010}.

While tabulating these opacity sources we identified several line
lists with more up-to-date line lists or opacity sources recommended
for use by the ExoMol project. These are: \methane
\citep[recommended:][]{yurchenko:2024ch4}, CN
\citep[recommended:][]{syme:2020,brooke:2014}, NO
\citep[recommended:][]{qu:2021}, OH \citep[recommended:][]{mitev:2025},
and SH \citep[recommended:][]{gorman:2019}.

\begin{deluxetable}{l l}[bt]
\tabletypesize{\scriptsize}
\tablecaption{  Opacity sources: \label{tab:opacity}}
\tablewidth{1.0\columnwidth}
\tablehead{
\colhead{Molecule} & \colhead{Reference}
}
\startdata
CH & \cite{masseron:2014} \\
CH$_3$ & \cite{adam:2019} \\
CH$_4$ & \cite{yurchenko:2014} \\  
C$_2$H$_2$ & \cite{chubb:2020} \\
C$_2$H$_4$ & \cite{mant:2018} \\
CN & \cite{brooke:2014} \\ 
CO & \cite{li:2015} \\
CO$_2$ & \cite{yurchenko:2020} \\
CS & \cite{paulose:2015} \\
HCN & \cite{barber:2014} \\
H$_2$CO & \cite{alrefaie:2015} \\
H$_2$O & \cite{polyansky:2018} \\
H$_2$O$_2$ & \cite{alrefaie:2016} \\
H$_2$S & \cite{azzam:2016} \\
NH & \citeauthor{brooke:2015} (\citeyear{brooke:2014nh,brooke:2015}) \\
NH$_3$ & \cite{coles:2019} \\
NO & \cite{rothman:2010}  \\ 
N$_2$O & \cite{yurchenko:2024n2o} \\
NS & \cite{yurchenko:2018} \\
OH & \cite{brooke:2016}   \\
SH & \cite{yurchenko:2018}  \\
SO & \cite{brady:2024} \\
SO$_2$ & \cite{underwood:2016} \\
\enddata
\end{deluxetable}



\appendix
The work presented here is similar in nature to that presented
recently by \cite{mukherjee:2025}.  Both efforts use grids of forward
models to explore the dependence of atmospheric chemistry in
H$_2$/He-dominated atmospheres on various parameters.  Here we list
some of the more salient differences in the two studies' modeling
approaches.

\begin{my_itemize}
  \item Modeled system: we use HAT-P-26b as our fiducial model, while
    they use WASP-39b.
  \item Temperature range: we span from 250--2050~K, while they span a
    narrower range of 400--1600~K.
  \item Internal temperature: we test only three values (100\,K,
    300\,K, 500\,K), while they test  five values from 30--500\,K.
  \item Temperature profiles: we calculate our profiles at a range of
    metallicities, from 0.3--1000$\times$ Solar abundances; they
    generate their profiles assuming the same abundance in all cases,
    namely 10$\times$ Solar.
  \item Vertical diffusion ($K_\mathrm{zz}$): We explore just three
    values of \kzz: $10^5, 10^7$, and $10^9$~cm$^2$~s$^{-1}$, with the
    middle value our nominal case; in contrast, they explore eight
    values even separated in log-space from $10^6$ to
    $10^{13}$~cm$^2$~s$^{-1}$ and use $10^9$~cm$^2$~s$^{-1}$ as the
    nominal, base case.
  \item Photochemistry: we run our photochemical models to steady
    state throughout the entire modeled atmosphere; they use a hybrid
    modeling approach assuming deep chemical equilibrium but with
    photochemistry dominating at higher altitudes.
  \item XUV flux: we explore a range of {\referee high-energy} irradiation with three
    stellar spectra spanning three orders of magnitude in XUV flux,
    and that flux is further scaled based on the modeled semimajor
    axis; \cite{mukherjee:2025} assume a single stellar XUV spectrum
    and do not scale the XUV flux incident on the planet with
    semimajor axis.
  \item Both studies also use different, independent modeling
    toolkits. We use \texttt{HELIOS} \citep{malik:2017} for our
    temperature profiles, \texttt{VULCAN} \citep{tsai:2017} for our
    photochemistry, and \texttt{petitRADTRANS} \citep{molliere:2019}
    to generate synthetic spectra; \cite{mukherjee:2025} use
    \texttt{PICASO} \citep{batalha:2019b}, \texttt{PhotoChem}
    \citep{wogan:2023}, and \texttt{PICASO}, respectively for those
    tasks.
\item Finally, that study is broader in scope, while here we provide a
  deeper exploration focused specifically on atmospheric sulfur
  chemistry.
\end{my_itemize}

\bibliographystyle{aasjournal}

\begin{thebibliography}{}
\expandafter\ifx\csname natexlab\endcsname\relax\def\natexlab#1{#1}\fi
\providecommand{\url}[1]{\href{#1}{#1}}
\providecommand{\dodoi}[1]{doi:~\href{http://doi.org/#1}{\nolinkurl{#1}}}
\providecommand{\doeprint}[1]{\href{http://ascl.net/#1}{\nolinkurl{http://ascl.net/#1}}}
\providecommand{\doarXiv}[1]{\href{https://arxiv.org/abs/#1}{\nolinkurl{https://arxiv.org/abs/#1}}}

\bibitem[{{Adam} {et~al.}(2019){Adam}, {Yachmenev}, {Yurchenko}, \&
  {Jensen}}]{adam:2019}
{Adam}, A.~Y., {Yachmenev}, A., {Yurchenko}, S.~N., \& {Jensen}, P. 2019,
  Journal of Physical Chemistry A, 123, 4755, \dodoi{10.1021/acs.jpca.9b02919}

\bibitem[{{Ahrer} {et~al.}(2025{\natexlab{a}}){Ahrer}, {Radica},
  {Piaulet-Ghorayeb}, {Raul}, {Wiser}, {Welbanks}, {Acuna}, {Allart},
  {Coulombe}, {Louca}, {MacDonald}, {Saidel}, {Evans-Soma}, {Benneke},
  {Christie}, {Beatty}, {Cadieux}, {Cloutier}, {Doyon}, {Fortney}, {Gagnebin},
  {Gapp}, {Innes}, {Knutson}, {Komacek}, {Krissansen-Totton}, {Miguel},
  {Pierrehumbert}, {Roy}, \& {Schlichting}}]{ahrer:2025}
{Ahrer}, E.-M., {Radica}, M., {Piaulet-Ghorayeb}, C., {et~al.}
  2025{\natexlab{a}}, arXiv e-prints, arXiv:2504.20428,
  \dodoi{10.48550/arXiv.2504.20428}

\bibitem[{{Ahrer} {et~al.}(2025{\natexlab{b}}){Ahrer}, {Gandhi}, {Alderson},
  {Kirk}, {Teske}, {Booth}, {McDonald}, {Christie}, {Claringbold}, {Nealon},
  {Panwar}, {Veras}, {Wakeford}, {Wheatley}, \& {Zamyatina}}]{ahrer:2025b}
{Ahrer}, E.-M., {Gandhi}, S., {Alderson}, L., {et~al.} 2025{\natexlab{b}},
  \mnras, \dodoi{10.1093/mnras/staf819}

\bibitem[{{Al-Refaie} {et~al.}(2016){Al-Refaie}, {Polyansky}, {Ovsyannikov},
  {Tennyson}, \& {Yurchenko}}]{alrefaie:2016}
{Al-Refaie}, A.~F., {Polyansky}, O.~L., {Ovsyannikov}, R.~I., {Tennyson}, J.,
  \& {Yurchenko}, S.~N. 2016, \mnras, 461, 1012, \dodoi{10.1093/mnras/stw1295}

\bibitem[{{Al-Refaie} {et~al.}(2015){Al-Refaie}, {Yachmenev}, {Tennyson}, \&
  {Yurchenko}}]{alrefaie:2015}
{Al-Refaie}, A.~F., {Yachmenev}, A., {Tennyson}, J., \& {Yurchenko}, S.~N.
  2015, \mnras, 448, 1704, \dodoi{10.1093/mnras/stv091}

\bibitem[{{Alderson} {et~al.}(2023){Alderson}, {Wakeford}, {Alam}, {Batalha},
  {Lothringer}, {Adams Redai}, {Barat}, {Brande}, {Damiano}, {Daylan},
  {Espinoza}, {Flagg}, {Goyal}, {Grant}, {Hu}, {Inglis}, {Lee}, {Mikal-Evans},
  {Ramos-Rosado}, {Roy}, {Wallack}, {Batalha}, {Bean}, {Benneke},
  {Berta-Thompson}, {Carter}, {Changeat}, {Col{\'o}n}, {Crossfield},
  {D{\'e}sert}, {Foreman-Mackey}, {Gibson}, {Kreidberg}, {Line},
  {L{\'o}pez-Morales}, {Molaverdikhani}, {Moran}, {Morello}, {Moses},
  {Mukherjee}, {Schlawin}, {Sing}, {Stevenson}, {Taylor}, {Aggarwal}, {Ahrer},
  {Allen}, {Barstow}, {Bell}, {Blecic}, {Casewell}, {Chubb}, {Crouzet},
  {Cubillos}, {Decin}, {Feinstein}, {Fortney}, {Harrington}, {Heng}, {Iro},
  {Kempton}, {Kirk}, {Knutson}, {Krick}, {Leconte}, {Lendl}, {MacDonald},
  {Mancini}, {Mansfield}, {May}, {Mayne}, {Miguel}, {Nikolov}, {Ohno}, {Palle},
  {Parmentier}, {Petit dit de la Roche}, {Piaulet}, {Powell}, {Rackham},
  {Redfield}, {Rogers}, {Rustamkulov}, {Tan}, {Tremblin}, {Tsai}, {Turner}, {de
  Val-Borro}, {Venot}, {Welbanks}, {Wheatley}, \& {Zhang}}]{alderson:2023}
{Alderson}, L., {Wakeford}, H.~R., {Alam}, M.~K., {et~al.} 2023, \nat, 614,
  664, \dodoi{10.1038/s41586-022-05591-3}

\bibitem[{{Arcangeli} {et~al.}(2018){Arcangeli}, {D{\'e}sert}, {Line}, {Bean},
  {Parmentier}, {Stevenson}, {Kreidberg}, {Fortney}, {Mansfield}, \&
  {Showman}}]{arcangeli:2018}
{Arcangeli}, J., {D{\'e}sert}, J.-M., {Line}, M.~R., {et~al.} 2018, \apjl, 855,
  L30, \dodoi{10.3847/2041-8213/aab272}

\bibitem[{{Azzam} {et~al.}(2016){Azzam}, {Tennyson}, {Yurchenko}, \&
  {Naumenko}}]{azzam:2016}
{Azzam}, A. A.~A., {Tennyson}, J., {Yurchenko}, S.~N., \& {Naumenko}, O.~V.
  2016, \mnras, 460, 4063, \dodoi{10.1093/mnras/stw1133}

\bibitem[{{Banerjee} {et~al.}(2024){Banerjee}, {Barstow}, {Gressier},
  {Espinoza}, {Sing}, {Allen}, {Birkmann}, {Challener}, {Crouzet}, {Haswell},
  {Lewis}, {Lewis}, \& {Yang}}]{bannerjee:2024}
{Banerjee}, A., {Barstow}, J.~K., {Gressier}, A., {et~al.} 2024, \apjl, 975,
  L11, \dodoi{10.3847/2041-8213/ad73d0}

\bibitem[{{Barber} {et~al.}(2014){Barber}, {Strange}, {Hill}, {Polyansky},
  {Mellau}, {Yurchenko}, \& {Tennyson}}]{barber:2014}
{Barber}, R.~J., {Strange}, J.~K., {Hill}, C., {et~al.} 2014, \mnras, 437,
  1828, \dodoi{10.1093/mnras/stt2011}

\bibitem[{{Batalha} {et~al.}(2019){Batalha}, {Marley}, {Lewis}, \&
  {Fortney}}]{batalha:2019b}
{Batalha}, N.~E., {Marley}, M.~S., {Lewis}, N.~K., \& {Fortney}, J.~J. 2019,
  \apj, 878, 70, \dodoi{10.3847/1538-4357/ab1b51}

\bibitem[{{Beatty} {et~al.}(2024){Beatty}, {Welbanks}, {Schlawin}, {Bell},
  {Line}, {Murphy}, {Edelman}, {Greene}, {Fortney}, {Henry}, {Mukherjee},
  {Ohno}, {Parmentier}, {Rauscher}, {Wiser}, \& {Arnold}}]{beatty:2024}
{Beatty}, T.~G., {Welbanks}, L., {Schlawin}, E., {et~al.} 2024, \apjl, 970,
  L10, \dodoi{10.3847/2041-8213/ad55e9}

\bibitem[{{Behr} {et~al.}(2023){Behr}, {France}, {Brown}, {Duvvuri}, {Bean},
  {Berta-Thompson}, {Froning}, {Miguel}, {Pineda}, {Wilson}, \&
  {Youngblood}}]{behr:2023}
{Behr}, P.~R., {France}, K., {Brown}, A., {et~al.} 2023, \aj, 166, 35,
  \dodoi{10.3847/1538-3881/acdb70}

\bibitem[{{Benneke} {et~al.}(2024){Benneke}, {Roy}, {Coulombe}, {Radica},
  {Piaulet}, {Ahrer}, {Pierrehumbert}, {Krissansen-Totton}, {Schlichting},
  {Hu}, {Yang}, {Christie}, {Thorngren}, {Young}, {Pelletier}, {Knutson},
  {Miguel}, {Evans-Soma}, {Dorn}, {Gagnebin}, {Fortney}, {Komacek},
  {MacDonald}, {Raul}, {Cloutier}, {Acuna}, {Lafreni{\`e}re}, {Cadieux},
  {Doyon}, {Welbanks}, \& {Allart}}]{benneke:2024}
{Benneke}, B., {Roy}, P.-A., {Coulombe}, L.-P., {et~al.} 2024, arXiv e-prints,
  arXiv:2403.03325, \dodoi{10.48550/arXiv.2403.03325}

\bibitem[{{Bergin} {et~al.}(2015){Bergin}, {Blake}, {Ciesla}, {Hirschmann}, \&
  {Li}}]{bergin:2015}
{Bergin}, E.~A., {Blake}, G.~A., {Ciesla}, F., {Hirschmann}, M.~M., \& {Li}, J.
  2015, Proceedings of the National Academy of Science, 112, 8965,
  \dodoi{10.1073/pnas.1500954112}

\bibitem[{{Bernath}(2020)}]{bernath:2020}
{Bernath}, P.~F. 2020, \jqsrt, 240, 106687, \dodoi{10.1016/j.jqsrt.2019.106687}

\bibitem[{{Binkert} \& {Birnstiel}(2023)}]{binkert:2023}
{Binkert}, F., \& {Birnstiel}, T. 2023, \mnras, 520, 2055,
  \dodoi{10.1093/mnras/stad182}

\bibitem[{{Brady} {et~al.}(2024){Brady}, {Yurchenko}, {Tennyson}, \&
  {Kim}}]{brady:2024}
{Brady}, R.~P., {Yurchenko}, S.~N., {Tennyson}, J., \& {Kim}, G.-S. 2024,
  \mnras, 527, 6675, \dodoi{10.1093/mnras/stad3508}

\bibitem[{Brooke {et~al.}(2016)Brooke, Bernath, Western, Sneden, Afşar, Li, \&
  Gordon}]{brooke:2016}
Brooke, J.~S., Bernath, P.~F., Western, C.~M., {et~al.} 2016, Journal of
  Quantitative Spectroscopy and Radiative Transfer, 168, 142,
  \dodoi{https://doi.org/10.1016/j.jqsrt.2015.07.021}

\bibitem[{{Brooke} {et~al.}(2015){Brooke}, {Bernath}, \&
  {Western}}]{brooke:2015}
{Brooke}, J. S.~A., {Bernath}, P.~F., \& {Western}, C.~M. 2015, \jcp, 143,
  026101, \dodoi{10.1063/1.4923422}

\bibitem[{Brooke {et~al.}(2014)Brooke, Bernath, Western, van Hemert, \&
  Groenenboom}]{brooke:2014nh}
Brooke, J. S.~A., Bernath, P.~F., Western, C.~M., van Hemert, M.~C., \&
  Groenenboom, G.~C. 2014, The Journal of Chemical Physics, 141, 054310,
  \dodoi{10.1063/1.4891468}

\bibitem[{{Brooke} {et~al.}(2014){Brooke}, {Ram}, {Western}, {Li}, {Schwenke},
  \& {Bernath}}]{brooke:2014}
{Brooke}, J. S.~A., {Ram}, R.~S., {Western}, C.~M., {et~al.} 2014, \apjs, 210,
  23, \dodoi{10.1088/0067-0049/210/2/23}

\bibitem[{{Chubb} {et~al.}(2020){Chubb}, {Tennyson}, \&
  {Yurchenko}}]{chubb:2020}
{Chubb}, K.~L., {Tennyson}, J., \& {Yurchenko}, S.~N. 2020, \mnras, 493, 1531,
  \dodoi{10.1093/mnras/staa229}

\bibitem[{{Coles} {et~al.}(2019){Coles}, {Yurchenko}, \&
  {Tennyson}}]{coles:2019}
{Coles}, P.~A., {Yurchenko}, S.~N., \& {Tennyson}, J. 2019, \mnras, 490, 4638,
  \dodoi{10.1093/mnras/stz2778}

\bibitem[{{Constantinou} {et~al.}(2023){Constantinou}, {Madhusudhan}, \&
  {Gandhi}}]{constantinou:2023}
{Constantinou}, S., {Madhusudhan}, N., \& {Gandhi}, S. 2023, \apjl, 943, L10,
  \dodoi{10.3847/2041-8213/acaead}

\bibitem[{{Crossfield}(2023)}]{crossfield:2023}
{Crossfield}, I. J.~M. 2023, \apjl, 952, L18, \dodoi{10.3847/2041-8213/ace35f}

\bibitem[{{Davenport} {et~al.}(2025){Davenport}, {Kempton}, {Nixon}, {Ih},
  {Deming}, {Fu}, {May}, {Bean}, {Gao}, {Rogers}, \& {Malik}}]{davenport:2025}
{Davenport}, B., {Kempton}, E. M.~R., {Nixon}, M.~C., {et~al.} 2025, \apjl,
  984, L44, \dodoi{10.3847/2041-8213/adcd76}

\bibitem[{{de Gruijter} {et~al.}(2025){de Gruijter}, {Tsai}, {Min}, {Waters},
  {Konings}, \& {Decin}}]{degruijter:2025}
{de Gruijter}, W., {Tsai}, S.-M., {Min}, M., {et~al.} 2025, \aap, 693, A132,
  \dodoi{10.1051/0004-6361/202450598}

\bibitem[{{Dyrek} {et~al.}(2024){Dyrek}, {Min}, {Decin}, {Bouwman}, {Crouzet},
  {Molli{\`e}re}, {Lagage}, {Konings}, {Tremblin}, {G{\"u}del}, {Pye},
  {Waters}, {Henning}, {Vandenbussche}, {Ardevol Martinez}, {Argyriou},
  {Ducrot}, {Heinke}, {van Looveren}, {Absil}, {Barrado}, {Baudoz},
  {Boccaletti}, {Cossou}, {Coulais}, {Edwards}, {Gastaud}, {Glasse}, {Glauser},
  {Greene}, {Kendrew}, {Krause}, {Lahuis}, {Mueller}, {Olofsson}, {Patapis},
  {Rouan}, {Royer}, {Scheithauer}, {Waldmann}, {Whiteford}, {Colina}, {van
  Dishoeck}, {{\"O}stlin}, {Ray}, \& {Wright}}]{dyrek:2024}
{Dyrek}, A., {Min}, M., {Decin}, L., {et~al.} 2024, \nat, 625, 51,
  \dodoi{10.1038/s41586-023-06849-0}

\bibitem[{{Feinstein} {et~al.}(2025){Feinstein}, {Booth}, {Bergner},
  {Lothringer}, {Matthews}, {Welbanks}, {Miguel}, {Bitsch}, {Eriksson}, {Kirk},
  {Pelletier}, {Penzlin}, {Piette}, {Piaulet-Ghorayeb}, {Schwarz}, {Turrini},
  {Acu{\~n}a-Aguirre}, {Ahrer}, {Barber}, {Brande}, {Chakrabarty},
  {Crossfield}, {Marleau}, {Huang}, {Johansen}, {Kreidberg}, {Livingston},
  {Luque}, {Oreshenko}, {Pacetti}, {Perotti}, {Polman}, {Prinoth}, {Semenov},
  {Simon}, {Teske}, \& {Whiteford}}]{feinstein:2025}
{Feinstein}, A.~D., {Booth}, R.~A., {Bergner}, J.~B., {et~al.} 2025, arXiv
  e-prints, arXiv:2506.00669, \dodoi{10.48550/arXiv.2506.00669}

\bibitem[{{Felix} {et~al.}(2025){Felix}, {Kitzmann}, {Demory}, \&
  {Mordasini}}]{felix:2025}
{Felix}, L., {Kitzmann}, D., {Demory}, B.-O., \& {Mordasini}, C. 2025, arXiv
  e-prints, arXiv:2504.13039, \dodoi{10.48550/arXiv.2504.13039}

\bibitem[{{Fortney}(2012)}]{fortney:2012}
{Fortney}, J.~J. 2012, \apjl, 747, L27, \dodoi{10.1088/2041-8205/747/2/L27}

\bibitem[{{France} {et~al.}(2016){France}, {Parke Loyd}, {Youngblood}, {Brown},
  {Schneider}, {Hawley}, {Froning}, {Linsky}, {Roberge}, {Buccino},
  {Davenport}, {Fontenla}, {Kaltenegger}, {Kowalski}, {Mauas}, {Miguel},
  {Redfield}, {Rugheimer}, {Tian}, {Vieytes}, {Walkowicz}, \&
  {Weisenburger}}]{france:2016}
{France}, K., {Parke Loyd}, R.~O., {Youngblood}, A., {et~al.} 2016, \apj, 820,
  89, \dodoi{10.3847/0004-637X/820/2/89}

\bibitem[{{Fu} {et~al.}(2024){Fu}, {Welbanks}, {Deming}, {Inglis}, {Zhang},
  {Lothringer}, {Ih}, {Moses}, {Schlawin}, {Knutson}, {Henry}, {Greene},
  {Sing}, {Savel}, {Kempton}, {Louie}, {Line}, \& {Nixon}}]{fu:2024}
{Fu}, G., {Welbanks}, L., {Deming}, D., {et~al.} 2024, \nat, 632, 752,
  \dodoi{10.1038/s41586-024-07760-y}

\bibitem[{{Gao} {et~al.}(2017){Gao}, {Marley}, {Zahnle}, {Robinson}, \&
  {Lewis}}]{gao:2017}
{Gao}, P., {Marley}, M.~S., {Zahnle}, K., {Robinson}, T.~D., \& {Lewis}, N.~K.
  2017, \aj, 153, 139, \dodoi{10.3847/1538-3881/aa5fab}

\bibitem[{{Gorman} {et~al.}(2019){Gorman}, {Yurchenko}, \&
  {Tennyson}}]{gorman:2019}
{Gorman}, M.~N., {Yurchenko}, S.~N., \& {Tennyson}, J. 2019, \mnras, 490, 1652,
  \dodoi{10.1093/mnras/stz2517}

\bibitem[{{Gressier} {et~al.}(2025){Gressier}, {Batalha}, {Wogan}, {Alderson},
  {Doud}, {Espinoza}, {MacDonald}, {Wakeford}, {Valenti}, {Lewis}, {Seager},
  {Stevenson}, {Allen}, {Ca{\~n}as}, {Challener}, {Glidden}, {Huang}, {Lin},
  {Louie}, {Maguire}, {Mullens}, {Sotzen}, {Valentine}, {Clampin}, {Pueyo},
  {van der Marel}, \& {Mountain}}]{gressier:2025}
{Gressier}, A., {Batalha}, N.~E., {Wogan}, N., {et~al.} 2025, \aj,
  arXiv:2509.16082.
\newblock \doarXiv{2509.16082}

\bibitem[{{Inglis} {et~al.}(2024){Inglis}, {Batalha}, {Lewis}, {Kataria},
  {Knutson}, {Kilpatrick}, {Gagnebin}, {Mukherjee}, {Pettyjohn}, {Crossfield},
  {Foote}, {Grant}, {Henry}, {Lally}, {McKemmish}, {Sing}, {Wakeford}, {Zapata
  Trujillo}, \& {Zellem}}]{inglis:2024}
{Inglis}, J., {Batalha}, N.~E., {Lewis}, N.~K., {et~al.} 2024, \apjl, 973, L41,
  \dodoi{10.3847/2041-8213/ad725e}

\bibitem[{{Khorshid} {et~al.}(2024){Khorshid}, {Min}, {Polman}, \&
  {Waters}}]{khorshid:2024}
{Khorshid}, N., {Min}, M., {Polman}, J., \& {Waters}, L.~B.~F.~M. 2024, \aap,
  685, A64, \dodoi{10.1051/0004-6361/202347124}

\bibitem[{{Kirk} {et~al.}(2024){Kirk}, {Ahrer}, {Claringbold}, {Zamyatina},
  {Fisher}, {McCormack}, {Panwar}, {Powell}, {Taylor}, {Thorngren}, {Christie},
  {Esparza-Borges}, {Tsai}, {Alderson}, {Booth}, {Fairman},
  {L{\'o}pez-Morales}, {Mayne}, {Meech}, {Molliere}, {Owen}, {Penzlin},
  {Sergeev}, {Valentine}, {Wakeford}, \& {Wheatley}}]{kirk:2024}
{Kirk}, J., {Ahrer}, E.-M., {Claringbold}, A.~B., {et~al.} 2024, arXiv
  e-prints, arXiv:2410.08116.
\newblock \doarXiv{2410.08116}

\bibitem[{{Kress} {et~al.}(2010){Kress}, {Tielens}, \&
  {Frenklach}}]{kress:2010}
{Kress}, M.~E., {Tielens}, A. G.~G.~M., \& {Frenklach}, M. 2010, Advances in
  Space Research, 46, 44, \dodoi{10.1016/j.asr.2010.02.004}

\bibitem[{{Lee} {et~al.}(2023){Lee}, {Tsai}, {Hammond}, \& {Tan}}]{lee:2023}
{Lee}, E. K.~H., {Tsai}, S.-M., {Hammond}, M., \& {Tan}, X. 2023, \aap, 672,
  A110, \dodoi{10.1051/0004-6361/202245473}

\bibitem[{{Lewis} {et~al.}(2013){Lewis}, {Knutson}, {Showman}, {Cowan},
  {Laughlin}, {Burrows}, {Deming}, {Crepp}, {Mighell}, {Agol}, {Bakos},
  {Charbonneau}, {D{\'e}sert}, {Fischer}, {Fortney}, {Hartman}, {Hinkley},
  {Howard}, {Johnson}, {Kao}, {Langton}, {Marcy}, \& {Winn}}]{lewis:2013}
{Lewis}, N.~K., {Knutson}, H.~A., {Showman}, A.~P., {et~al.} 2013, ArXiv
  e-prints.
\newblock \doarXiv{1302.5084}

\bibitem[{Li {et~al.}(2015)Li, Gordon, Rothman, Tan, Hu, Kassi, Campargue, \&
  Medvedev}]{li:2015}
Li, G., Gordon, I.~E., Rothman, L.~S., {et~al.} 2015, The Astrophysical Journal
  Supplement Series, 216, 15.
\newblock \url{http://stacks.iop.org/0067-0049/216/i=1/a=15}

\bibitem[{{Lodders}(2003)}]{lodders:2003}
{Lodders}, K. 2003, \apj, 591, 1220, \dodoi{10.1086/375492}

\bibitem[{{Lodders}(2020)}]{lodders:2020}
---. 2020, Oxford Research Enc. of Pl. Sci.,
  \dodoi{10.1093/acrefore/9780190647926.013.145}

\bibitem[{{Lothringer} {et~al.}(2018){Lothringer}, {Barman}, \&
  {Koskinen}}]{lothringer:2018b}
{Lothringer}, J.~D., {Barman}, T., \& {Koskinen}, T. 2018, \apj, 866, 27,
  \dodoi{10.3847/1538-4357/aadd9e}

\bibitem[{{Lothringer} {et~al.}(2021){Lothringer}, {Rustamkulov}, {Sing},
  {Gibson}, {Wilson}, \& {Schlaufman}}]{lothringer:2021}
{Lothringer}, J.~D., {Rustamkulov}, Z., {Sing}, D.~K., {et~al.} 2021, \apj,
  914, 12, \dodoi{10.3847/1538-4357/abf8a9}

\bibitem[{{Ma} {et~al.}(2025){Ma}, {Saba}, {Faris Al-Refaie}, {Tinetti},
  {Yurchenko}, {Tennyson}, \& {Cecchi Pestellini}}]{ma:2025}
{Ma}, S., {Saba}, A., {Faris Al-Refaie}, A., {et~al.} 2025, arXiv e-prints,
  arXiv:2504.07823, \dodoi{10.48550/arXiv.2504.07823}

\bibitem[{{Malik} {et~al.}(2017){Malik}, {Grosheintz}, {Mendon{\c{c}}a},
  {Grimm}, {Lavie}, {Kitzmann}, {Tsai}, {Burrows}, {Kreidberg}, {Bedell},
  {Bean}, {Stevenson}, \& {Heng}}]{malik:2017}
{Malik}, M., {Grosheintz}, L., {Mendon{\c{c}}a}, J.~M., {et~al.} 2017, \aj,
  153, 56, \dodoi{10.3847/1538-3881/153/2/56}

\bibitem[{{Mant} {et~al.}(2018){Mant}, {Yachmenev}, {Tennyson}, \&
  {Yurchenko}}]{mant:2018}
{Mant}, B.~P., {Yachmenev}, A., {Tennyson}, J., \& {Yurchenko}, S.~N. 2018,
  \mnras, 478, 3220, \dodoi{10.1093/mnras/sty1239}

\bibitem[{{Masseron} {et~al.}(2014){Masseron}, {Plez}, {Van Eck}, {Colin},
  {Daoutidis}, {Godefroid}, {Coheur}, {Bernath}, {Jorissen}, \&
  {Christlieb}}]{masseron:2014}
{Masseron}, T., {Plez}, B., {Van Eck}, S., {et~al.} 2014, \aap, 571, A47,
  \dodoi{10.1051/0004-6361/201423956}

\bibitem[{{Mayo} {et~al.}(2024){Mayo}, {Fortenbach}, {Louie}, {Dressing},
  {Turtelboom}, {Giacalone}, \& {Harada}}]{mayo:2025}
{Mayo}, A.~W., {Fortenbach}, C.~D., {Louie}, D.~R., {et~al.} 2024, arXiv
  e-prints, arXiv:2501.00609, \dodoi{10.48550/arXiv.2501.00609}

\bibitem[{{Meech} {et~al.}(2025){Meech}, {Claringbold}, {Ahrer}, {Kirk},
  {L{\'o}pez-Morales}, {Taylor}, {Booth}, {Penzlin}, {Alderson}, {Christie},
  {Esparza-Borges}, {Fairman}, {Mayne}, {McCormack}, {Owen}, {Panwar},
  {Powell}, {Sergeev}, {Valentine}, {Wakeford}, {Wheatley}, \&
  {Zamyatina}}]{meech:2025}
{Meech}, A., {Claringbold}, A.~B., {Ahrer}, E.-M., {et~al.} 2025, \mnras, 539,
  1381, \dodoi{10.1093/mnras/staf530}

\bibitem[{{Mitev} {et~al.}(2025){Mitev}, {Bowesman}, {Zhang}, {Yurchenko}, \&
  {Tennyson}}]{mitev:2025}
{Mitev}, G.~B., {Bowesman}, C.~A., {Zhang}, J., {Yurchenko}, S.~N., \&
  {Tennyson}, J. 2025, \mnras, 536, 3401, \dodoi{10.1093/mnras/stae2803}

\bibitem[{{Molli{\`e}re} {et~al.}(2019){Molli{\`e}re}, {Wardenier}, {van
  Boekel}, {Henning}, {Molaverdikhani}, \& {Snellen}}]{molliere:2019}
{Molli{\`e}re}, P., {Wardenier}, J.~P., {van Boekel}, R., {et~al.} 2019, \aap,
  627, A67, \dodoi{10.1051/0004-6361/201935470}

\bibitem[{{Molli{\`e}re} {et~al.}(2022){Molli{\`e}re}, {Molyarova}, {Bitsch},
  {Henning}, {Schneider}, {Kreidberg}, {Eistrup}, {Burn}, {Nasedkin},
  {Semenov}, {Mordasini}, {Schlecker}, {Schwarz}, {Lacour}, {Nowak}, \&
  {Schulik}}]{molliere:2022}
{Molli{\`e}re}, P., {Molyarova}, T., {Bitsch}, B., {et~al.} 2022, \apj, 934,
  74, \dodoi{10.3847/1538-4357/ac6a56}

\bibitem[{{Mordasini} {et~al.}(2016){Mordasini}, {van Boekel}, {Molli{\`e}re},
  {Henning}, \& {Benneke}}]{mordasini:2016}
{Mordasini}, C., {van Boekel}, R., {Molli{\`e}re}, P., {Henning}, T., \&
  {Benneke}, B. 2016, \apj, 832, 41, \dodoi{10.3847/0004-637X/832/1/41}

\bibitem[{{Moses} {et~al.}(2024){Moses}, {Tsai}, {Fortney}, {Constantinou},
  {Madhusudhan}, {Visscher}, {Yu}, {Plane}, {Yang}, {Zahnle}, \&
  {Lee}}]{moses:2024}
{Moses}, J., {Tsai}, S.-M., {Fortney}, J., {et~al.} 2024, in AAS/Division for
  Planetary Sciences Meeting Abstracts, Vol.~56, 56th Annual Meeting of the
  Division for Planetary Sciences, 308.06

\bibitem[{{Mukherjee} {et~al.}(2024){Mukherjee}, {Fortney}, {Wogan}, {Sing}, \&
  {Ohno}}]{mukherjee:2025}
{Mukherjee}, S., {Fortney}, J.~J., {Wogan}, N.~F., {Sing}, D.~K., \& {Ohno}, K.
  2024, arXiv e-prints, arXiv:2410.17169, \dodoi{10.48550/arXiv.2410.17169}

\bibitem[{{Pacetti} {et~al.}(2022){Pacetti}, {Turrini}, {Schisano}, {Molinari},
  {Fonte}, {Politi}, {Hennebelle}, {Klessen}, {Testi}, \&
  {Lebreuilly}}]{pacetti:2022}
{Pacetti}, E., {Turrini}, D., {Schisano}, E., {et~al.} 2022, \apj, 937, 36,
  \dodoi{10.3847/1538-4357/ac8b11}

\bibitem[{{Paulose} {et~al.}(2015){Paulose}, {Barton}, {Yurchenko}, \&
  {Tennyson}}]{paulose:2015}
{Paulose}, G., {Barton}, E.~J., {Yurchenko}, S.~N., \& {Tennyson}, J. 2015,
  \mnras, 454, 1931, \dodoi{10.1093/mnras/stv1543}

\bibitem[{{Polman} {et~al.}(2023){Polman}, {Waters}, {Min}, {Miguel}, \&
  {Khorshid}}]{polman:2023}
{Polman}, J., {Waters}, L.~B.~F.~M., {Min}, M., {Miguel}, Y., \& {Khorshid}, N.
  2023, \aap, 670, A161, \dodoi{10.1051/0004-6361/202244647}

\bibitem[{{Polyansky} {et~al.}(2018){Polyansky}, {Kyuberis}, {Zobov},
  {Tennyson}, {Yurchenko}, \& {Lodi}}]{polyansky:2018}
{Polyansky}, O.~L., {Kyuberis}, A.~A., {Zobov}, N.~F., {et~al.} 2018, \mnras,
  480, 2597, \dodoi{10.1093/mnras/sty1877}

\bibitem[{{Powell} {et~al.}(2024){Powell}, {Feinstein}, {Lee}, {Zhang}, {Tsai},
  {Taylor}, {Kirk}, {Bell}, {Barstow}, {Gao}, {Bean}, {Blecic}, {Chubb},
  {Crossfield}, {Jordan}, {Kitzmann}, {Moran}, {Morello}, {Moses}, {Welbanks},
  {Yang}, {Zhang}, {Ahrer}, {Bello-Arufe}, {Brande}, {Casewell}, {Crouzet},
  {Cubillos}, {Demory}, {Dyrek}, {Flagg}, {Hu}, {Inglis}, {Jones}, {Kreidberg},
  {L{\'o}pez-Morales}, {Lagage}, {Meier Vald{\'e}s}, {Miguel}, {Parmentier},
  {Piette}, {Rackham}, {Radica}, {Redfield}, {Stevenson}, {Wakeford},
  {Aggarwal}, {Alam}, {Batalha}, {Batalha}, {Benneke}, {Berta-Thompson},
  {Brady}, {Caceres}, {Carter}, {D{\'e}sert}, {Harrington}, {Iro}, {Line},
  {Lothringer}, {MacDonald}, {Mancini}, {Molaverdikhani}, {Mukherjee}, {Nixon},
  {Oza}, {Palle}, {Rustamkulov}, {Sing}, {Steinrueck}, {Venot}, {Wheatley}, \&
  {Yurchenko}}]{powell:2024}
{Powell}, D., {Feinstein}, A.~D., {Lee}, E. K.~H., {et~al.} 2024, \nat, 626,
  979, \dodoi{10.1038/s41586-024-07040-9}

\bibitem[{{Prinoth} {et~al.}(2022){Prinoth}, {Hoeijmakers}, {Kitzmann},
  {Sandvik}, {Seidel}, {Lendl}, {Borsato}, {Thorsbro}, {Anderson}, {Barrado},
  {Kravchenko}, {Allart}, {Bourrier}, {Cegla}, {Ehrenreich}, {Fisher}, {Lovis},
  {Guzm{\'a}n-Mesa}, {Grimm}, {Hooton}, {Morris}, {Oreshenko}, {Pino}, \&
  {Heng}}]{prinoth:2022}
{Prinoth}, B., {Hoeijmakers}, H.~J., {Kitzmann}, D., {et~al.} 2022, Nature
  Astronomy, 6, 449, \dodoi{10.1038/s41550-021-01581-z}

\bibitem[{{Qu} {et~al.}(2021){Qu}, {Yurchenko}, \& {Tennyson}}]{qu:2021}
{Qu}, Q., {Yurchenko}, S.~N., \& {Tennyson}, J. 2021, \mnras, 504, 5768,
  \dodoi{10.1093/mnras/stab1154}

\bibitem[{{Rothman} {et~al.}(2010){Rothman}, {Gordon}, {Barber}, {Dothe},
  {Gamache}, {Goldman}, {Perevalov}, {Tashkun}, \& {Tennyson}}]{rothman:2010}
{Rothman}, L., {Gordon}, I., {Barber}, R., {et~al.} 2010, JQSRT, 111, 2139,
  \dodoi{10.1016/j.jqsrt.2010.05.001}

\bibitem[{{Rustamkulov} {et~al.}(2023){Rustamkulov}, {Sing}, {Mukherjee},
  {May}, {Kirk}, {Schlawin}, {Line}, {Piaulet}, {Carter}, {Batalha}, {Goyal},
  {L{\'o}pez-Morales}, {Lothringer}, {MacDonald}, {Moran}, {Stevenson},
  {Wakeford}, {Espinoza}, {Bean}, {Batalha}, {Benneke}, {Berta-Thompson},
  {Crossfield}, {Gao}, {Kreidberg}, {Powell}, {Cubillos}, {Gibson}, {Leconte},
  {Molaverdikhani}, {Nikolov}, {Parmentier}, {Roy}, {Taylor}, {Turner},
  {Wheatley}, {Aggarwal}, {Ahrer}, {Alam}, {Alderson}, {Allen}, {Banerjee},
  {Barat}, {Barrado}, {Barstow}, {Bell}, {Blecic}, {Brande}, {Casewell},
  {Changeat}, {Chubb}, {Crouzet}, {Daylan}, {Decin}, {D{\'e}sert},
  {Mikal-Evans}, {Feinstein}, {Flagg}, {Fortney}, {Harrington}, {Heng}, {Hong},
  {Hu}, {Iro}, {Kataria}, {Kempton}, {Krick}, {Lendl}, {Lillo-Box}, {Louca},
  {Lustig-Yaeger}, {Mancini}, {Mansfield}, {Mayne}, {Miguel}, {Morello},
  {Ohno}, {Palle}, {Petit dit de la Roche}, {Rackham}, {Radica},
  {Ramos-Rosado}, {Redfield}, {Rogers}, {Shkolnik}, {Southworth}, {Teske},
  {Tremblin}, {Tucker}, {Venot}, {Waalkes}, {Welbanks}, {Zhang}, \&
  {Zieba}}]{rustamkulov:2023}
{Rustamkulov}, Z., {Sing}, D.~K., {Mukherjee}, S., {et~al.} 2023, \nat, 614,
  659, \dodoi{10.1038/s41586-022-05677-y}

\bibitem[{{Schneider} \& {Bitsch}(2021)}]{schneider:2021b}
{Schneider}, A.~D., \& {Bitsch}, B. 2021, \aap, 654, A72,
  \dodoi{10.1051/0004-6361/202141096}

\bibitem[{{Sing} {et~al.}(2024){Sing}, {Rustamkulov}, {Thorngren}, {Barstow},
  {Tremblin}, {Alves de Oliveira}, {Beck}, {Birkmann}, {Challener}, {Crouzet},
  {Espinoza}, {Ferruit}, {Giardino}, {Gressier}, {Lee}, {Lewis}, {Maiolino},
  {Manjavacas}, {Rauscher}, {Sirianni}, \& {Valenti}}]{sing:2024}
{Sing}, D.~K., {Rustamkulov}, Z., {Thorngren}, D.~P., {et~al.} 2024, arXiv
  e-prints, arXiv:2405.11027, \dodoi{10.48550/arXiv.2405.11027}

\bibitem[{{Syme} \& {McKemmish}(2020)}]{syme:2020}
{Syme}, A.-M., \& {McKemmish}, L.~K. 2020, \mnras, 499, 25,
  \dodoi{10.1093/mnras/staa2791}

\bibitem[{{Thao} {et~al.}(2024){Thao}, {Mann}, {Feinstein}, {Gao}, {Thorngren},
  {Rotman}, {Welbanks}, {Brown}, {Duvvuri}, {France}, {Longo}, {Sandoval},
  {Schneider}, {Wilson}, {Youngblood}, {Vanderburg}, {Barber}, {Wood},
  {Batalha}, {Kraus}, {Murray}, {Newton}, {Rizzuto}, {Tofflemire}, {Tsai},
  {Bean}, {Berta-Thompson}, {Evans-Soma}, {Froning}, {Kempton}, {Miguel}, \&
  {Pineda}}]{thao:2024}
{Thao}, P.~C., {Mann}, A.~W., {Feinstein}, A.~D., {et~al.} 2024, arXiv
  e-prints, arXiv:2409.16355, \dodoi{10.48550/arXiv.2409.16355}

\bibitem[{{Tsai} {et~al.}(2017){Tsai}, {Lyons}, {Grosheintz}, {Rimmer},
  {Kitzmann}, \& {Heng}}]{tsai:2017}
{Tsai}, S.-M., {Lyons}, J.~R., {Grosheintz}, L., {et~al.} 2017, \apjs, 228, 20,
  \dodoi{10.3847/1538-4365/228/2/20}

\bibitem[{{Tsai} {et~al.}(2023{\natexlab{a}}){Tsai}, {Moses}, {Powell}, \&
  {Lee}}]{tsai:2023b}
{Tsai}, S.-M., {Moses}, J.~I., {Powell}, D., \& {Lee}, E. K.~H.
  2023{\natexlab{a}}, \apjl, 959, L30, \dodoi{10.3847/2041-8213/ad1405}

\bibitem[{{Tsai} {et~al.}(2023{\natexlab{b}}){Tsai}, {Steinrueck},
  {Parmentier}, {Lewis}, \& {Pierrehumbert}}]{tsai:2023c}
{Tsai}, S.-M., {Steinrueck}, M., {Parmentier}, V., {Lewis}, N., \&
  {Pierrehumbert}, R. 2023{\natexlab{b}}, \mnras, 520, 3867,
  \dodoi{10.1093/mnras/stad214}

\bibitem[{{Tsai} {et~al.}(2023{\natexlab{c}}){Tsai}, {Lee}, {Powell}, {Gao},
  {Zhang}, {Moses}, {H{\'e}brard}, {Venot}, {Parmentier}, {Jordan}, {Hu},
  {Alam}, {Alderson}, {Batalha}, {Bean}, {Benneke}, {Bierson}, {Brady},
  {Carone}, {Carter}, {Chubb}, {Inglis}, {Leconte}, {Line},
  {L{\'o}pez-Morales}, {Miguel}, {Molaverdikhani}, {Rustamkulov}, {Sing},
  {Stevenson}, {Wakeford}, {Yang}, {Aggarwal}, {Baeyens}, {Barat}, {de
  Val-Borro}, {Daylan}, {Fortney}, {France}, {Goyal}, {Grant}, {Kirk},
  {Kreidberg}, {Louca}, {Moran}, {Mukherjee}, {Nasedkin}, {Ohno}, {Rackham},
  {Redfield}, {Taylor}, {Tremblin}, {Visscher}, {Wallack}, {Welbanks},
  {Youngblood}, {Ahrer}, {Batalha}, {Behr}, {Berta-Thompson}, {Blecic},
  {Casewell}, {Crossfield}, {Crouzet}, {Cubillos}, {Decin}, {D{\'e}sert},
  {Feinstein}, {Gibson}, {Harrington}, {Heng}, {Henning}, {Kempton}, {Krick},
  {Lagage}, {Lendl}, {Lothringer}, {Mansfield}, {Mayne}, {Mikal-Evans},
  {Palle}, {Schlawin}, {Shorttle}, {Wheatley}, \& {Yurchenko}}]{tsai:2023}
{Tsai}, S.-M., {Lee}, E. K.~H., {Powell}, D., {et~al.} 2023{\natexlab{c}},
  \nat, 617, 483, \dodoi{10.1038/s41586-023-05902-2}

\bibitem[{{Tsai} {et~al.}(2024){Tsai}, {Parmentier}, {Mendon{\c{c}}a}, {Tan},
  {Deitrick}, {Hammond}, {Savel}, {Zhang}, {Pierrehumbert}, \&
  {Schwieterman}}]{tsai:2024}
{Tsai}, S.-M., {Parmentier}, V., {Mendon{\c{c}}a}, J.~M., {et~al.} 2024, \apj,
  963, 41, \dodoi{10.3847/1538-4357/ad1600}

\bibitem[{{Turrini} {et~al.}(2021){Turrini}, {Schisano}, {Fonte}, {Molinari},
  {Politi}, {Fedele}, {Pani{\'c}}, {Kama}, {Changeat}, \&
  {Tinetti}}]{turrini:2021}
{Turrini}, D., {Schisano}, E., {Fonte}, S., {et~al.} 2021, \apj, 909, 40,
  \dodoi{10.3847/1538-4357/abd6e5}

\bibitem[{Underwood {et~al.}(2016)Underwood, Tennyson, Yurchenko, Huang,
  Schwenke, Lee, Clausen, \& Fateev}]{underwood:2016}
Underwood, D.~S., Tennyson, J., Yurchenko, S.~N., {et~al.} 2016, Monthly
  Notices of the Royal Astronomical Society, 459, 3890,
  \dodoi{10.1093/mnras/stw849}

\bibitem[{{Valenti} {et~al.}(2025){Valenti}, {Lewis}, {Gressier}, {MacDonald},
  {Alderson}, {Challener}, {Wakeford}, \& {Espinoza}}]{valenti:2025}
{Valenti}, J., {Lewis}, N., {Gressier}, A., {et~al.} 2025, in ExoClimes,
  Vol.~1, ExoClimes VII, Poster 9

\bibitem[{{Veillet} {et~al.}(2025){Veillet}, {Venot}, {Sirjean}, {Citrangolo
  Destro}, {Fournet}, {Al-Refaie}, {H{\'e}brard}, {Glaude}, \&
  {Bounaceur}}]{veillet:2025}
{Veillet}, R., {Venot}, O., {Sirjean}, B., {et~al.} 2025, arXiv e-prints,
  arXiv:2505.12152, \dodoi{10.48550/arXiv.2505.12152}

\bibitem[{{Wakeford} {et~al.}(2017){Wakeford}, {Sing}, {Kataria}, {Deming},
  {Nikolov}, {Lopez}, {Tremblin}, {Amundsen}, {Lewis}, {Mandell}, {Fortney},
  {Knutson}, {Benneke}, \& {Evans}}]{wakeford:2017}
{Wakeford}, H.~R., {Sing}, D.~K., {Kataria}, T., {et~al.} 2017, Science, 356,
  628, \dodoi{10.1126/science.aah4668}

\bibitem[{{Welbanks} {et~al.}(2019){Welbanks}, {Madhusudhan}, {Allard},
  {Hubeny}, {Spiegelman}, \& {Leininger}}]{welbanks:2019}
{Welbanks}, L., {Madhusudhan}, N., {Allard}, N.~F., {et~al.} 2019, \apjl, 887,
  L20, \dodoi{10.3847/2041-8213/ab5a89}

\bibitem[{{Welbanks} {et~al.}(2024){Welbanks}, {Bell}, {Beatty}, {Line},
  {Ohno}, {Fortney}, {Schlawin}, {Greene}, {Rauscher}, {McGill}, {Murphy},
  {Parmentier}, {Tang}, {Edelman}, {Mukherjee}, {Wiser}, {Lagage}, {Dyrek}, \&
  {Arnold}}]{welbanks:2024}
{Welbanks}, L., {Bell}, T.~J., {Beatty}, T.~G., {et~al.} 2024, \nat, 630, 836,
  \dodoi{10.1038/s41586-024-07514-w}

\bibitem[{{Wogan}(2023)}]{wogan:2023}
{Wogan}, N. 2023, {PhotochemPy: 1-D photochemical model of rocky planet
  atmospheres}, Astrophysics Source Code Library, record ascl:2312.011

\bibitem[{{Youngblood} {et~al.}(2016){Youngblood}, {France}, {Loyd}, {Linsky},
  {Redfield}, {Schneider}, {Wood}, {Brown}, {Froning}, {Miguel}, {Rugheimer},
  \& {Walkowicz}}]{youngblood:2016}
{Youngblood}, A., {France}, K., {Loyd}, R.~O.~P., {et~al.} 2016, \apj, 824,
  101, \dodoi{10.3847/0004-637X/824/2/101}

\bibitem[{{Yurchenko} {et~al.}(2018){Yurchenko}, {Bond}, {Gorman}, {Lodi},
  {McKemmish}, {Nunn}, {Shah}, \& {Tennyson}}]{yurchenko:2018}
{Yurchenko}, S.~N., {Bond}, W., {Gorman}, M.~N., {et~al.} 2018, \mnras, 478,
  270, \dodoi{10.1093/mnras/sty939}

\bibitem[{{Yurchenko} {et~al.}(2020){Yurchenko}, {Mellor}, {Freedman}, \&
  {Tennyson}}]{yurchenko:2020}
{Yurchenko}, S.~N., {Mellor}, T.~M., {Freedman}, R.~S., \& {Tennyson}, J. 2020,
  \mnras, 496, 5282, \dodoi{10.1093/mnras/staa1874}

\bibitem[{{Yurchenko} {et~al.}(2024{\natexlab{a}}){Yurchenko}, {Mellor}, \&
  {Tennyson}}]{yurchenko:2024n2o}
{Yurchenko}, S.~N., {Mellor}, T.~M., \& {Tennyson}, J. 2024{\natexlab{a}},
  \mnras, 534, 1364, \dodoi{10.1093/mnras/stae2201}

\bibitem[{{Yurchenko} {et~al.}(2024{\natexlab{b}}){Yurchenko}, {Owens},
  {Kefala}, \& {Tennyson}}]{yurchenko:2024ch4}
{Yurchenko}, S.~N., {Owens}, A., {Kefala}, K., \& {Tennyson}, J.
  2024{\natexlab{b}}, \mnras, 528, 3719, \dodoi{10.1093/mnras/stae148}

\bibitem[{{Yurchenko} \& {Tennyson}(2014)}]{yurchenko:2014}
{Yurchenko}, S.~N., \& {Tennyson}, J. 2014, \mnras, 440, 1649,
  \dodoi{10.1093/mnras/stu326}

\bibitem[{{Zahnle} {et~al.}(2009){Zahnle}, {Marley}, \&
  {Fortney}}]{zahnle:2009}
{Zahnle}, K., {Marley}, M.~S., \& {Fortney}, J.~J. 2009, ArXiv e-prints.
\newblock \doarXiv{0911.0728}

\bibitem[{{Zhang} {et~al.}(2025){Zhang}, {Paragas}, {Bean}, {Yeung}, {Chachan},
  {Greene}, {Lunine}, \& {Deming}}]{zhang:2025}
{Zhang}, M., {Paragas}, K., {Bean}, J.~L., {et~al.} 2025, \aj, 169, 38,
  \dodoi{10.3847/1538-3881/ad8cd2}

\end{thebibliography}

\end{document}